\documentclass[useAMS,usenatbib]{mn2e}
\usepackage{times}
\usepackage{graphicx,epsfig}

\title[Feedback and ICM enrichment in galaxy groups]{Temperature and
  abundance profiles of hot gas in galaxy groups --\\ II. Implications
  for feedback and ICM enrichment}

\author[J. Rasmussen and T. J. Ponman]{Jesper
    Rasmussen$^{1}$\thanks{E-mail: jr@ociw.edu}\thanks{Chandra Fellow}
    and Trevor J.~Ponman$^{2}$ \\ $^{1}$Carnegie Observatories, 813
    Santa Barbara Street, Pasadena, California 91101, USA\\ $^{2}$
    School of Physics and Astronomy, University of Birmingham,
    Edgbaston, Birmingham B15 2TT}
\begin{document} 
 
\date{} 
 
\pagerange{\pageref{firstpage}--\pageref{lastpage}} \pubyear{2008} 
 
\maketitle 
 
\label{firstpage} 
 
\begin{abstract}
  We investigate the history of galactic feedback and chemical
  enrichment within a sample of 15 X-ray bright groups of galaxies, on
  the basis of the inferred Fe and Si distributions in the hot gas and
  the associated metal masses produced by core-collapse and type~Ia
  supernovae (SN). Most of these cool-core groups show a central Fe
  and Si excess, which can be explained by prolonged enrichment by
  SN~Ia and stellar winds in the central early-type galaxy alone, but
  with tentative evidence for additional processes contributing to
  core enrichment in hotter groups. Inferred metal mass-to-light
  ratios inside $r_{500}$ show a positive correlation with total group
  mass but are generally significantly lower than in clusters, due to
  a combination of lower global ICM abundances and gas-to-light ratios
  in groups. This metal deficiency is present for products from both
  SN~Ia and SN~II, and suggests that metals were either synthesized,
  released from galaxies, or retained within the ICM less efficiently
  in lower-mass systems. We explore possible causes, including
  variations in galaxy formation and metal release efficiency,
  cooling-out of metals, and gas and metal loss via AGN-- or
  starburst-driven galactic winds from groups or their precursor
  filaments. Loss of enriched material from filaments coupled with
  post-collapse AGN feedback emerge as viable explanations, but we
  also find evidence for metals to have been released less efficiently
  from galaxies in cooler groups and for the ICM in these to appear
  chemically less evolved, possibly reflecting more extended star
  formation histories in less massive systems.  Some implications for
  the hierarchical growth of clusters from groups are briefly
  discussed.

\end{abstract} 
 
 
\section{Introduction}

Groups of galaxies are repositories for a substantial fraction of all
baryons and metals at low redshift. Being much more numerous than more
massive clusters and more susceptible to the effects of
non-gravitational processes such as galaxy winds driven by supernovae
or active galactic nuclei (e.g., \citealt{hels00,borg04,rome06}), they
represent unique laboratories for the study of galactic feedback and
chemical enrichment of the intergalactic medium. In practice, this can
be exploited through X-ray observations of the hot intragroup gas,
which allow the abundance of several heavy elements in this gas to be
constrained.

Numerous such studies have been undertaken for massive galaxy
clusters, providing a reasonably well-constrained picture of the
amount and distribution of metals in the intracluster medium (ICM) and
in the cluster galaxies themselves \citep*{fino00, boeh04, degr04,
mouh06, bald07, lecc08}.  However, despite extensive observational and
theoretical efforts, several important questions related to ICM
enrichment still remain, such as the relative importance of the
mechanisms responsible for transferring metals from galaxies to the
ICM, and the role of mergers and galactic wind activity in
redistributing enriched material.  For example, it has been
notoriously difficult to reproduce observed ICM metal distributions in
cosmological hydrodynamical simulations of clusters, suggesting that
important processes affecting baryons in these systems may still be
inadequately captured or altogether missing in such models (see, e.g.,
review by \citealt{borg08b}). Characterization of the chemical makeup
of the hot gas in more numerous lower-mass groups could provide an
important ingredient in resolving these questions. Due to the
structural role played by groups in connecting the `field' -- i.e.\
isolated galaxies and the intergalactic medium -- with the rare
high-density peaks represented by massive clusters, a clearer picture
of enrichment in groups may have ramifications for our understanding
of the evolution of baryons across a wide range of environments.

In an early study of the distribution of elemental abundances in 
intragroup gas, \citet{fino99} investigated three groups, for which they 
found significant spatial variations in the ratio of iron to
$\alpha$--element abundances, concluding that both supernovae (SNe)
type~Ia and II must have contributed to the enrichment of the
intracluster medium in groups. Evidence for significant iron gradients
in a sample of 10 groups and early-type galaxies were also found from
{\em ROSAT} data by \citet{buot00b}. \citet*{fino01} followed this up
using a larger sample of 18 cool clusters, but with only two systems
at $T<2$~keV, while \citet{fino02} studied nine groups mainly in the
context of their entropy profiles and the constraints they imposed on
preheating scenarios. More recent work include the two-dimensional
{\em XMM-Newton} results of \citet{fino06,fino07}, but the main focus
here was the entropy and pressure structure of the ICM.  Thus, while
most of these studies have focused on $T\la 2$~keV systems, detailed
discussions of the abundance distributions and the implications for
galaxy and group evolution have generally been limited and based on
samples of just a few groups (e.g., \citealt{fino99}). Hence, a
dedicated study of chemical enrichment in a substantial number of
groups has been lacking.

In the first paper in this series \citep[hereafter Paper~I]{rasm07},
we presented projected radial profiles of temperature and Fe and Si
abundance of the hot intragroup gas in a sample of 15 groups,
discussing the results in a statistical framework. The groups were
selected mainly from the GEMS sample of \citet{osmo04}, on the basis
of their good photon statistics in available {\em Chandra} data and
for their relatively undisturbed X-ray morphology. The low {\em
Chandra} background enabled robust abundance constraints to reasonably
large radii, with the majority of our groups having Fe and Si profiles
extracted to at least $\frac{2}{3}r_{500}$, where $r_{500}$ is the
radius enclosing a mean density of 500 times the critical density.  At
the same time, the spatial resolution of {\em Chandra} allowed
sampling of group cores at a level of detail usually unattainable with
other X-ray telescopes.  With only one exception (NGC\,4125), all the
groups were found to contain cooler gas in the central regions. Main
results of the study included the discovery that Fe abundances in the
groups show a highly significant decrease with radius, declining
continually from a central excess towards a value at $r_{500}$ which
is, on average, lower than corresponding values seen in more massive
clusters by a factor of two.  In addition, it was found that Fe
enrichment in group cores is dominated by SN~Ia as also found for more
massive clusters (e.g., \citealt{degr04}), thus confirming early,
tentative {\em ASCA} results \citep{fino00}. The SN~II contribution to
enrichment increases with radius, however, and potentially becomes
exclusive at radii approaching $r_{500}$.

The purpose of the present paper is to explore some of the
implications of these results. In Section~\ref{sec,est} we outline the
derivation of gas and metal masses and the optical properties of the
groups.  Section~\ref{sec,results} presents the main results inferred
from these quantities and from the observed abundance distributions,
including the metal masses produced by SN~Ia and SN~II, the
corresponding mass-to-light ratios, and the impact of SN in the
brightest group galaxy (BGG) and in the groups as a whole. One finding
is that the metal mass-to-light ratios deviate substantially from
results for more massive systems, and we explore a variety of possible
explanations for this in Section~\ref{sec,fate}, along with some
further implications for the feedback and chemical enrichment history
of the groups in Section~\ref{sec,filaments}.
Section~\ref{sec,conclude} briefly discusses and summarizes the
results, including some general implications for structure formation.

As in Paper~I, we use the solar abundance table of \citet{grev98},
theoretical supernova yields from \citet{iwam99} and \citet{nomo06},
and the Kendall rank-order correlation coefficient $\tau$ to quantify
the significance $\sigma_K$ of linear relationships.  A value of
$H_0=70$~km~s$^{-1}$~Mpc$^{-1}$ is assumed throughout, and all errors
are given at the 68~per~cent confidence level unless otherwise stated.

\section{Metal masses and optical luminosities within $r_{500}$}\label{sec,est}

Results from Paper~I for the abundance profiles are summarised in
Fig.~\ref{fig,norm_bin}.
\begin{figure*} 
 \includegraphics[width=176mm]{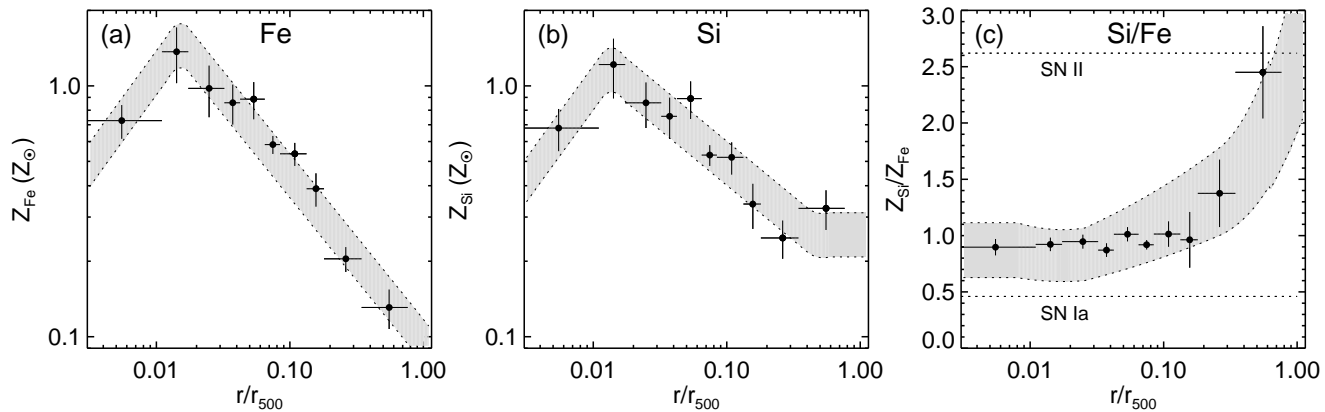} 
 \caption{Fe and Si abundance profiles for all groups, in radial bins
   of 20 data points, along with the corresponding Si/Fe
   ratios. Shaded areas illustrate the adopted $1\sigma$ error
   envelopes on our parametrized abundance profiles. Dotted lines in
   (c) show the expectations for pure SN~Ia and SN~II enrichment.}
\label{fig,norm_bin} 
\end{figure*} 
The data points here are similar to those of fig.~11 in Paper~I,
showing the results in radial bins of 20 measurements each. We briefly
recapitulate the distinct features of these profiles, which are (i) a
central abundance excess for both Fe and Si, often accompanied by a
dip in the very centre for both elements; (ii) a clear decline in Fe
abundance with radius towards a value of $\sim 0.1$~Z$_\odot$ at
$r_{500}$, with a somewhat less pronounced decline for Si (which
furthermore flattens out or even rises at large radii). For reference,
the SN yields adopted here, as explained in Paper~I, are $M_{\rm
Fe}\approx 0.79$~M$_\odot$ and $M_{\rm Si}\approx 0.21$~M$_\odot$ for
Fe and Si from SN~Ia, and $M_{\rm Fe}\approx 0.08$~M$_\odot$ and
$M_{\rm Si}\approx 0.12$~M$_\odot$ for core-collapse supernovae
including SN~II. This implies $Z_{\rm Si}/Z_{\rm Fe}\approx
0.46$~Z$_{\rm Si,\odot}$/Z$_{\rm Fe,\odot}$ for SN~Ia ejecta,
2.62~Z$_{\rm Si,\odot}$/Z$_{\rm Fe,\odot}$ for those of SN~II, and
with a solar abundance ratio requiring 3.4~SN~II per SN~Ia.  The
impact of assuming different sets of SN yields will be discussed in
Section~\ref{sec,results}.

\subsection{Parametrizing the abundance profiles}

For the majority of our groups, abundances could not be measured all
the way out to $r_{500}$.  In order to enable predictions of $Z(r)$
beyond the radius of measurement for individual groups, as well as to
facilitate visualisation of sample-averaged results, we made
parametrizations of the average abundance profiles in
Fig.~\ref{fig,norm_bin}.  This approach can be justified on the
grounds that while the statistical uncertainties on $Z_{\rm Fe}$ and
$Z_{\rm Si}$ can be quite large for individual groups, the intrinsic
scatter in these quantities is reasonably small outside the innermost
regions. Outside the peak of the profiles in Fig.~\ref{fig,norm_bin},
we have adopted the fitted relations of Paper~I [equations~(8) and
(9)] for these parametrizations. Inside the peaks, a linear rise in
log\,$Z$ vs.\ log\,$(r/r_{500})$ has been assumed, normalised such as
to ensure a continuous behaviour at the peaks. Outside $r=0.1r_{500}$,
where the vast majority of gas resides, the typical standard error on
the mean in the individual bins in Fig.~\ref{fig,norm_bin}a and
\ref{fig,norm_bin}b is 16~per~cent for $Z_{\rm Fe}$ and 18~per~cent
for $Z_{\rm Si}$. We have conservatively adopted 20~per~cent relative
errors as representative for our parametrizations of both
profiles. These 20~per~cent error envelopes are shown as shaded
regions in Fig.~\ref{fig,norm_bin}; within these errors, the
parametrizations are consistent with every data point in the figure.
Note that the shaded region in Fig.~\ref{fig,norm_bin}c is not a
direct parametrization of the Si/Fe ratios, but results from the ratio
of the parametrized Si and Fe profiles, with their adopted errors
added in quadrature.

Unless otherwise stated, our analysis employs the observed abundance
profiles of individual groups out to the edge of the usable {\em
Chandra data}, extrapolating these to larger radii where necessary
using the above parametrizations and their associated errors. Though
obviously a model-dependent approach, the adopted parametrizations are
at least empirically motivated across all radii considered here, and
the resulting Si/Fe ratios remain consistent with commonly used SN
model yields everywhere inside $r_{500}$. Furthermore, as will be
demonstrated, there is no clear evidence for the observed abundance
profiles at large radii to be systematically different for, for
example, high-- and low--temperature groups, so the adopted
parametrizations should be reasonably representative for the entire
group sample.  The additional precaution of adopting conservative
errors on these was taken to further alleviate concerns regarding this
approach.

\subsection{Derivation of gas and metal masses}\label{sec,mgas}

For the calculation of gas and metal masses of the groups, we adopted
a standard 1-D $\beta$--model to describe the gas density profiles,
using values of $\beta$ and $r_c$ for each group taken from the
literature. These values are listed in Table~\ref{tab,gas} below,
taken from the {\em ROSAT} analyses of \citet{mulc03} for NGC\,507,
NGC\,4125, and NGC\,7619, \citet{davi96} for NGC\,2300, and
\citet{osmo04} for the remainder.

The gas density profile of each group was normalised by means of the
measured {\em Chandra} flux in the 0.1--0.3~$r_{500}$ range, using the
spectral normalisation $A$ from {\sc xspec},
\begin{equation}
  A = \frac{10^{-14}}{4\pi D^2 (1+z)^2} 
  \int_{0.1r_{500}}^{0.3r_{500}}n_en_{\rm H} \mbox{d}V  \mbox{ cm$^{-5}$}, 
\label{eq,xspec}
\end{equation} 
where $D$ is the group distance (from Paper~I) and $n_e$ and $n_{\rm
H}$ are the number densities of electrons and hydrogen atoms,
respectively. The reason for normalising the profiles using the flux
outside $r=0.1r_{500}$ is that many of the published gas density
parameters result from two-component fits, with one component used to
describe the emission in and around the central galaxy, and the other,
of interest here, to describe the extended group emission. This
approach also reduces the impact of projection effects on the
resulting normalisations, and suppresses effects related to blurring
by the {\em ROSAT} point spread function. Where relevant, a geometric
correction was applied to $A$, to account for lost detector area due
to removed point sources and incomplete azimuthal coverage of the
relevant radial range (see Paper~I for details). The median correction
factor was 1.38, with no significant dependence on, for example, mean
group temperature. Gas and metal masses were then calculated using the
adopted $\beta$--model parameters. Errors on these masses within the
radius of interest were obtained on the basis of 5,000 Monte Carlo
realisations of the gas density and abundance profiles of each
group. For each realisation, gas density parameters ($A$, $\beta$,
$r_c$) and abundances ($Z_{\rm Fe}$, $Z_{\rm Si}$) in each radial bin
were drawn from Gaussians with $\sigma$ equal to the associated
$1\sigma$ error.  The final metal mass and its uncertainty were then
obtained as the mean and $\sigma$ of the resulting distribution of
masses. These distributions are generally symmetric, as illustrated in
Fig.~\ref{fig,histo} which shows an example outcome of this approach
for the calculation of Fe masses.

\begin{figure} 
\mbox{\hspace{-1mm} 
 \includegraphics[width=84mm]{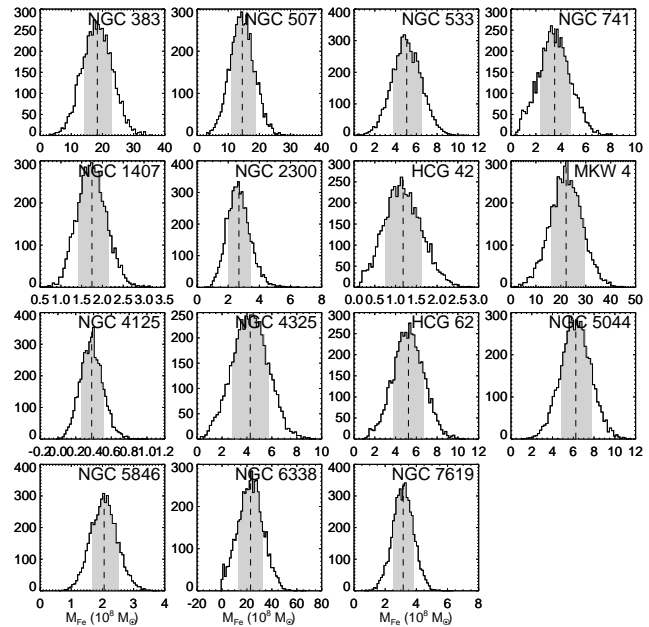}} 
\caption{Illustration of the Monte Carlo output for the calculation of
  gas and metal masses and their uncertainties.  Histograms show the
  obtained distribution of Fe masses for each group, with the
  resulting mean values represented by dashed lines, and the
  associated $\pm 1\sigma$ uncertainty ranges by the shaded area.}
\label{fig,histo} 
\end{figure} 

As is the case for the abundance profiles, the gas density profiles
have generally not been measured out to $r_{500}$ within our
sample. Where relevant, the adopted $\beta$-parameters were therefore
used to extrapolate the gas density profile of each group beyond the
radius of measurement. Systematic uncertainties related to this
extrapolation are not straightforward to quantify, because the number
of gas mass measurements extending to $r_{500}$ in $T\la 2$~keV
systems are still fairly limited (see, e.g., \citealt{sun09}). {\em
XMM-Newton} studies of groups to large radii \citep{rasm04} suggest
that standard $\beta$--models can provide a good description of the
profiles of relaxed groups out to, and beyond, $r_{500}$.  However,
several studies have found the slope of the gas density profile in
more massive clusters to steepen progressively at large radii, with
typical slopes at $r_{500}$ that are $\sim 15$~per~cent steeper than
the `canonical' cluster value of $\beta \approx 2/3$ (e.g.,
\citealt{vikh06,maug08}). To account for this steepening, Vikhlinin
et~al.\ (2006; see also \citealt{sun09}) multiply the standard
$\beta$--model for the density profile by a factor
$[(1+r^{\gamma}/r_s^{\gamma})^{\epsilon/\gamma}]^{-1}$, with
$\gamma=3$ and $r_s$ and $\epsilon$ fitted parameters.

Although the \citet{vikh06} study was largely limited to $\langle T
\rangle \ga 2$~keV systems (leaving it unclear to what extent these
results can be reliably extended to the regime of cool groups, whose
density profiles inside $r_{500}$ tend to be flatter than those of
clusters; \citealt{sand03b}), we have investigated the impact of
adopting the \citet{vikh06} parametrization for our groups. Mean
values for the Vikhlinin clusters are $r_s =1.1 r_{500}$ and
$\epsilon=3.2$ (from their table~2), with no systematic dependence of
these parameters on cluster temperature; adopting these values, we
find that our gas masses within $r_{500}$ could, on average, be
overestimated by 24~per~cent (with a small standard deviation of
4~per~cent). This is within the adopted systematic uncertainty on
$M_{\rm gas}$ (see Section~\ref{sec,comparison}), so within these
adopted errors, our gas masses should be robust towards such changes
in the assumed shape of the density profile. Furthermore, any trends
in $M_{\rm gas}$--based quantities with mean system temperature would
not be significantly affected by such a change. Many of the cluster
comparison results presented in the following may also be subject to
similar corrections, so our overall conclusions should remain valid
irrespective of such a change.

\subsection{Optical, near-infrared, and radio properties}\label{sec,optical}

In order to derive mass-to-light ratios and to estimate the expected
supernova rate in the groups, $K$- and $B$-band luminosities of the
group members were also derived inside $r_{500}$.  Luminosities were
computed from extinction-corrected galaxy magnitudes in NED using the
method of \citet{osmo04}: Galaxies were extracted inside $r_{500}$ and
within a velocity interval of $3\sigma_{\rm v}$ around the mean
recession velocity $v_r$ of the group, using starting values of $v_r$
and velocity dispersion $\sigma_{\rm v}$ from \citet{osmo04} where
available, or from \citet{mulc03} otherwise. These quantities were
then re-evaluated iteratively until convergence was reached.  The only
exception to this rule was NGC\,1407, for which a velocity range
extending to $3.6\sigma_{\rm v}$ was employed. This was done in order
to include the second brightest group galaxy, NGC\,1400, which has an
unusually large peculiar velocity of $\sim 1100$~km~s$^{-1}$ relative
to the group mean \citep*{tren06}.

The adopted $K$-band magnitudes of the group members generally derive
from Two-Micron All Sky Survey data (2MASS; \citealt{skru06}).
Inspection of a composite histogram of the apparent magnitudes
suggests that incompleteness sets in around $m_K \approx 12.5$ for our
groups, consistent with the expectation that the 2MASS source catalog
is 90~per~cent complete down to $m_K = 13.5$ for extended sources more
than $30^{\circ}$ from the Galactic plane \citep{skru06}, a region
covering 12 of our 15 groups.  For each group, we corrected for
incompleteness below the absolute magnitude corresponding to $m_K
=12.5$ using the Schechter fit of \citet*{lin04b} to the $K$-band
luminosity function of 93 groups and clusters (and including their
correction factor to obtain the Schechter normalisation inside
$r_{500}$ instead of $r_{200}$). The resulting corrections are
typically small, on average 5~per~cent.

In order to facilitate comparison to existing results in the
literature, $B$-band magnitudes of the group members were also
extracted from NED in a similar manner.  In four of the groups,
magnitudes had to be converted from the $R$-band for a few
galaxies. This was done assuming $B$--$R=1.5$ as is appropriate for
early-type galaxies. While this could lead to a slight underestimation
of $L_B$, we found that using instead $B-R=0.8$ as for spirals
resulted in a mere 4~per~cent increase in $L_B$ for NGC\,4325, with
the change being less than 0.5~per~cent for the remaining three
groups. We have refrained from taking these minute corrections into
account in the error budget.  Since the optical magnitudes generally
derive from a variety of heterogeneous data sets with varying
completeness, we have not attempted a similar correction for
incompleteness as applied to the $K$-band data.  Following
\citet{osmo04}, we instead simply aimed to homogenize completeness
across the sample by applying an absolute magnitude cut to all
groups. This cut discarded galaxies fainter than $M_B=-16.32$,
retaining an expected 90~per~cent of the $B$-band luminosity of the
galaxy population.

Finally, using the derived $B$- and $K$-band luminosities and the
corresponding $B$--$K$ colours, estimates of the total stellar masses
$M_\ast$ of the groups and their central galaxies were obtained using
the coefficients listed in table~1 of \citet{bell01}, which imply
\begin{equation}
  \mbox{log} \left(\frac{M_\ast/L_K}{\mbox{M$_\odot$/L$_\odot$}}\right) =
  0.212(B-K) -0.959
  \label{eq,mstar}
\end{equation}
for the $K$-band stellar mass-to-light ratio of the group members. We
note that our approach implies that our $B$--$K$ colours could be
slightly but systematically biased towards the red, because we are
attempting to account for all $K$-band light by performing a
completeness correction while only retaining an expected 90~per~cent
of the $B$-band light. This could lead to overestimates of stellar
mass which, while potentially varying from group to group, should not
exceed 0.02~dex, i.e.\ 5~per~cent, and so this bias has no substantive
impact on our results.

\subsection{Robustness of results}\label{sec,comparison}

Although not themselves the focus of the present study, reliable gas
masses are still important ingredients in this work, so we briefly
discuss the robustness of our adopted approach. For example, while the
derived statistical errors from our Monte Carlo calculations are small
for most groups, one concern is whether the adopted
$\beta$--parameters and our method of normalising the density profile
are robust. Another is that our approach does not take into account
radial variations of the plasma emissivity. To explore the potential
impact of this, we have compared our gas mass estimates to published
values for the groups for which we have been able to find useful
comparison data in the literature, consisting of either {\em Chandra},
{\em XMM}, or {\em Suzaku} measurements published within a given
radius using different approaches \citep{mori06,gast07,komi09}. At
larger radii, we have also compared our results within $r_{500}$ to
those obtained using the density parametrizations of \citet{sand03}
based on {\em ROSAT} data. This comparison is shown in
Fig.~\ref{fig,compare}.
\begin{figure} 
\mbox{\hspace{-1mm} 
 \includegraphics[width=83mm]{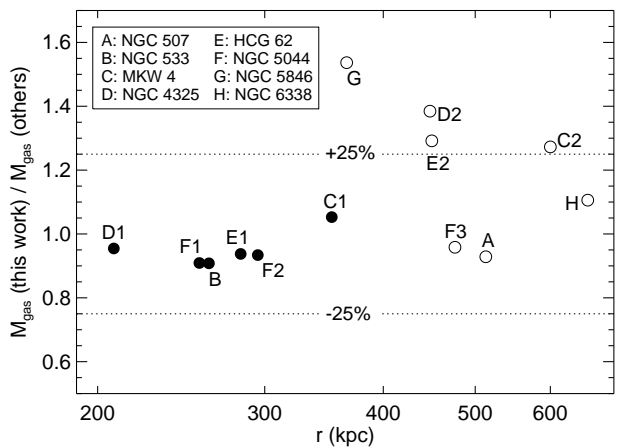}} 
\caption{Ratio of our estimated gas masses to other published
  results. Filled symbols are based on measured values from the
  literature, empty ones on our values within $r_{500}$
  (Table~\ref{tab,gas}) compared to those resulting from extrapolating
  the gas density parametrizations of \citet{sand03} to the same
  physical radius. Dotted lines mark a 25~per~cent difference from
  equality. Comparison values for $M_{\rm gas}$ are: {\em B}:
  $8.7\times 10^{11}$~M$_\odot$ within $r=262$~kpc \citep{gast07};
  {\em C1}: $2.8\times 10^{12}$~M$_\odot$ \citep[$r \le
  353$~kpc]{gast07}; {\em D1}: $6.6\times 10^{11}$~M$_\odot$ \citep[$r
  \le 208$~kpc]{gast07}; {\em E1}: $1.6\times 10^{12}$~M$_\odot$
  \citep[$r \le 283$~kpc]{mori06}; {\em F1}: $9.9\times
  10^{11}$~M$_\odot$ \citep[$r \le 256$~kpc]{komi09}; {\em F2}:
  $1.2\times 10^{12}$~M$_\odot$ \citep[$r \le 295$~kpc]{gast07}.}
\label{fig,compare} 
\end{figure} 

The figure indicates that there is no clear systematic variation
between our results and other estimates, with most estimates agreeing
to within $\sim 25$~per~cent. An exception is the clearly discrepant
NGC\,5846, for which the \citet{sand03} parametrization yields a gas
mass within $r_{500}$ which is only 65~per~cent of our
estimate. However, even when including this system, the average
deviation of our estimates from existing results is only 10~per~cent
for all measurements, or 17~per cent if only including the independent
largest-radius estimate for each group. We therefore consider a
25~per~cent error a reasonably conservative estimate of the systematic
uncertainty associated with our method of gas mass estimation, so we
have added this uncertainty in quadrature to those related to any
quantity based on $M_{\rm gas}$.

Since the time of our original {\em Chandra} analysis, performed using
{\sc ciao} 3.0 with calibration database CALDB 3.2.3, updated
calibration data including a recalibration of the {\em Chandra}
effective area have been released. To briefly test the potential
impact of this update on our results, we rederived the temperature and
abundance profiles for NGC\,4325, a group typical of our sample in
terms of signal-to-noise ratio and mean ICM temperature. For this
updated analysis, we employed the most recent calibration data, CALDB
4.1.2, with {\sc ciao} 4.1, following the spectral fitting procedure
described in Paper~I. We find that the updated temperature
measurements are perfectly consistent with the original values in all
radial bins, and the results for the Fe and Si profiles, shown in
Fig.~\ref{fig,N4325}, also demonstrate good consistency throughout,
with no systematic offset between the new and original results. In
fact, the main effect of using the updated calibration data is a
considerable reduction in the statistical uncertainties of most
measurements. We therefore conclude that our original results should
generally be robust to the recent changes to the {\em Chandra}
calibration.

\begin{figure} 
 \includegraphics[width=80mm]{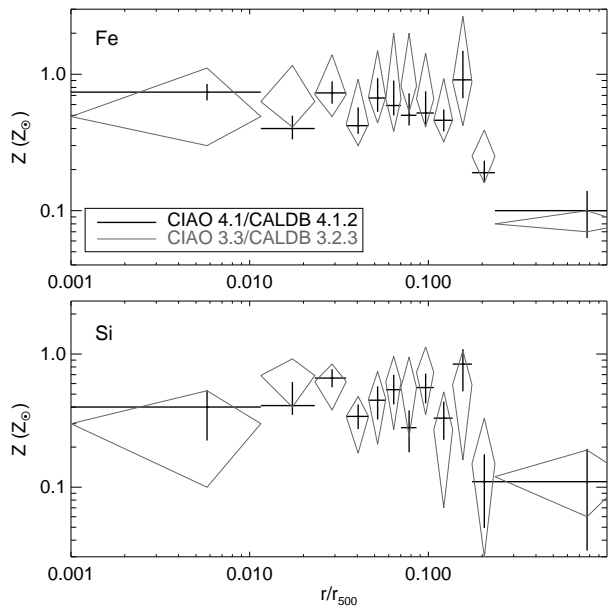}
 \caption{Comparison of (top) Fe and (bottom) Si profiles derived for
     NGC\,4325 using the most recent {\em Chandra} calibration data
     (black crosses) and those of our original analysis in Paper~I
     (gray diamonds).}
\label{fig,N4325}
\end{figure} 

As a further indicator of the robustness of our abundance
measurements, we point to the good overall agreement with the
independently obtained {\em XMM} results presented by
\citet{fino06,fino07} (see discussion in Paper~I), in particular the
excellent agreement across the full radial range between our
sample-averaged Fe profile in Fig.~\ref{fig,norm_bin} and the
corresponding {\em XMM} result (Johnson et~al., in
preparation). Encouragingly, good agreement is also seen when
comparing our metal masses derived within specific radii to the
corresponding {\em XMM} results for NGC\,5044 \citep*{buot04} and the
{\em Suzaku} results for HCG\,62 \citep{toko08}, NGC\,507
\citep{sato09}, and NGC\,5044 \citep{komi09}, once differences in the
adopted abundance table have been taken into account. This is
illustrated in Fig.~\ref{fig,compare2} (left panel), demonstrating
consistency at the $\sim 1\sigma$~level between ours and existing
results in all four cases. Comparing also our resulting Fe
mass-to-light ratios to the corresponding {\em Suzaku} estimates in
Fig.~\ref{fig,compare2} (right panel), good overall agreement is again
seen, with HCG\,62 as the only exception. Since the derived Fe masses
for this group are consistent, this discrepancy stems from differences
in the assumed value of $L_B$ within $r=13$~arcmin, with the value of
\citet{toko08}, $L_B=4.9\times 10^{10}$~L$_\odot$ (derived by
converting $R$-band magnitudes to the $B$-band assuming a fixed
$B$--$R$ colour), being significantly lower than our value of
$L_B=7.6\times 10^{10}$~L$_\odot$ obtained with the approach outlined
in the previous Section. This minor issue notwithstanding, we conclude
that our results for both gas masses, metal masses, and metal
mass-to-light ratios generally show good agreement with existing
results in the literature, thus lending credibility to our approach.

\begin{figure} 
 \includegraphics[width=84mm]{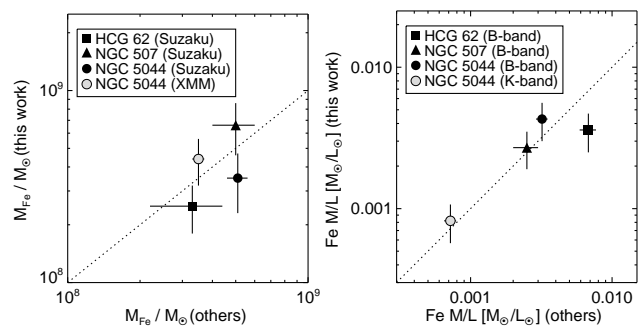}
 \caption{Comparison of (left) our Fe masses and (right) mass-to-light
      ratios to existing results, where available. Dotted lines
      represent equality.}
\label{fig,compare2}
\end{figure}

\section{Results}\label{sec,results}

This section summarizes the derived group properties that will form
the basis of the discussion in the following sections.
Table~\ref{tab,gas} lists the main group properties used in this
study, including the mean temperature $\langle T \rangle$ of each
group derived inside the range $0.1-0.3r_{500}$ (from Paper~I), the
adopted gas density parameters, derived metal masses inside $r_{500}$
and their fractional contribution from SN~Ia, the total $B$- and
$K$-band luminosities inside the same radius, and the derived galaxy
velocity dispersions. Some of the properties of the brightest group
galaxy (BGG) in each system (in all cases an early-type galaxy at the
centre of the X-ray emission) are also relevant to the following
discussion.  In addition to stellar masses derived using
equation~(\ref{eq,mstar}), we have also compiled central stellar
velocity dispersions $\sigma_{0,\ast}$ for these galaxies from
Hyperleda (available for all the BGGs except NGC\,4325 and 6338), and
the 1.4-GHz luminosity density of any central radio source, based on
fluxes listed in NED. These results are listed in Table~\ref{tab,BGG}.
\begin{table*} 
 \centering 
   \caption{Gas density parameters, gas and metal masses, optical and
     near-infrared luminosities, and galaxy velocity dispersions for
     the sample, all obtained inside $r_{500}$. Also listed is the
     corresponding fractional metal mass $f_{\rm Ia}$ provided by
     SN~Ia for both Fe and Si.}
  \label{tab,gas} 
  \begin{tabular}{@{}lccrcrcrcccc@{}} \hline 
   \multicolumn{1}{l}{Group} & 
\multicolumn{1}{c}{$\langle T_{\rm X} \rangle$} & 
\multicolumn{1}{c}{$\beta$} & 
\multicolumn{1}{c}{$r_c$} &
\multicolumn{1}{c}{$M_{\rm gas}$} &
\multicolumn{1}{c}{$M_{\rm Fe}$} &
\multicolumn{1}{c}{$f_{\rm Ia}$} &
\multicolumn{1}{c}{$M_{\rm Si}$} &
\multicolumn{1}{c}{$f_{\rm Ia}$} &
\multicolumn{1}{c}{$L_B$} &
\multicolumn{1}{c}{$L_K$} &
\multicolumn{1}{c}{$\sigma_{\rm v}$} \\
  &  (keV) &   & (kpc) \hspace{3mm} & ($10^{12}$~M$_\odot$) & 
  ($10^8$~M$_\odot$) & (Fe) & ($10^8$~M$_\odot$) & (Si) 
  & (log~L$_\odot$) & (log~L$_\odot$) &  (km~s$^{-1}$)  \\ \hline 
NGC\,383  & $1.65^{+0.04}_{-0.06}$ & $0.36\pm 0.00$ & $2.1\pm 0.2$  & 
    $5.4 \pm 0.1$ & $18.4\pm 4.7$ & 0.35 & $21.2\pm 7.9$ & 0.08 & 11.50 & 
    12.26 & 450 \\  
NGC\,507  & $1.30^{+0.03}_{-0.03}$ & $0.42\pm 0.01$ & $25.7\pm 0.6$ & 
    $3.9 \pm 0.2$ & $14.4\pm 3.9$ & 0.31 & $16.4\pm 6.3$ & 0.07 & 11.69 & 
    12.35 & 635 \\   
NGC\,533  & $1.22^{+0.05}_{-0.05}$ & $0.42\pm 0.01$ & $2.2\pm 1.7$  & 
    $2.3 \pm 0.1$ & $5.0\pm 1.4$ & 0.25 & $6.2\pm 2.7$   & 0.06 & 11.47 & 
    12.11 & 439 \\ 
NGC\,741  & $1.42^{+0.14}_{-0.12}$ & $0.44\pm 0.01$ & $2.3\pm 0.2$ & 
    $2.7 \pm 0.1$ & $3.5\pm 1.3$ & 0.34 & $11.9\pm 3.8$  & 0.03 & 11.35 & 
    12.03 & 453 \\ 
NGC\,1407 & $1.01^{+0.07}_{-0.09}$ & $0.46\pm 0.01$ & $0.1\pm 0.1$ & 
    $0.7 \pm 0.1$ & $1.7\pm 0.3$ & 0.31 & $2.0\pm 0.6$   & 0.07 & 11.04 & 
    11.73 &   319 \\ 
NGC\,2300 & $0.78^{+0.04}_{-0.03}$ & $0.41\pm 0.03$ & $56.5\pm 16.8$ & 
    $1.2 \pm 0.4$ & $2.8\pm 1.0$ & 0.19  & $3.8\pm 1.8$  & 0.04 & 10.86 & 
    11.49 &  300 \\ 
HCG\,42   & $0.80^{+0.05}_{-0.05}$ & $0.56\pm 0.02$ & $4.7\pm 0.7$ & 
    $0.6 \pm 0.1$ &  $1.2\pm 0.4$ & 0.38 & $1.3\pm 0.6$  & 0.10 & 11.32 & 
    11.92 &   282 \\ 
MKW\,4    & $1.78^{+0.07}_{-0.09}$ & $0.43\pm 0.01$ & $9.4\pm 2.9$ & 
    $7.0 \pm 0.2$ & $22.3\pm 6.6$ & 0.54 & $19.7\pm 9.7$ & 0.16 & 11.78 & 
    12.44 & 565 \\ 
NGC\,4125 & $0.33^{+0.12}_{-0.05}$ & $0.47\pm 0.03$ & $<0.7$ & 
    $0.2 \pm 0.1$ & $0.4\pm 0.1$ & 0.27 & $0.6\pm 0.3$   & 0.05 & 10.62 & 
    11.29 &   56 \\ 
NGC\,4325 & $0.99^{+0.02}_{-0.02}$ & $0.58\pm 0.01$ & $27.6\pm 5.0$ & 
    $1.8 \pm 0.1$ & $4.3\pm 1.4$ & 0.71 & $2.8\pm 1.5$   & 0.30 & 10.96 & 
    11.73 &   376 \\ 
HCG\,62   & $1.00^{+0.03}_{-0.03}$ & $0.48\pm 0.01$ & $2.4\pm 0.3$ & 
    $3.1 \pm 0.1$ & $5.3\pm 1.4$ & 0.52 &  $10.7\pm 3.5$ & 0.06 & 11.35 & 
    12.02 & 418 \\ 
NGC\,5044 & $1.12^{+0.03}_{-0.03}$ & $0.51\pm 0.00$ & $6.0\pm 0.2$ & 
    $2.3 \pm 0.1$ & $6.3\pm 1.5$ & 0.42 & $6.6\pm 2.0$   & 0.10 & 11.22 & 
    11.92 &   426 \\ 
NGC\,5846 & $0.66^{+0.04}_{-0.03}$ & $0.51\pm 0.01$ & $2.2\pm 0.3$ & 
    $0.9 \pm 0.1$ & $2.1\pm 0.4$ & 0.29 & $2.6\pm 0.8$   & 0.06 & 11.23 & 
    11.90 &   346 \\ 
NGC\,6338 & $2.13^{+0.19}_{-0.07}$ & $0.44\pm 0.01$ & $3.7\pm 1.0$ & 
    $9.4 \pm 0.3$ & $22.6\pm 10.2$ & 0.55 & $34.6\pm 14.7$ & 0.05 & 11.54 & 
    12.37 & 651 \\ 
NGC\,7619 & $1.06^{+0.07}_{-0.03}$ & $0.44\pm 0.01$ & $7.9\pm 0.5$ & 
    $1.5 \pm 0.1$ & $3.2\pm 0.7$ & 0.45 & $3.2\pm 1.2$ & 0.12 & 11.44 & 
    12.12 &  557 \\ 
\hline 
\end{tabular} 
\end{table*}

\begin{table} 
 \centering
   \caption{Summary of relevant properties of the brightest group
   galaxy, including the derived stellar mass, central stellar
   velocity dispersion, and 1.4-GHz radio power of any central radio
   source.}
  \label{tab,BGG} 
  \begin{tabular}{@{}lcccc@{}} \hline 
   \multicolumn{1}{l}{BGG} & 
\multicolumn{1}{c}{$L_K$} & 
\multicolumn{1}{c}{$M_\ast$} & 
\multicolumn{1}{c}{$\sigma_{0,\ast}$} &
\multicolumn{1}{c}{$P_{1.4}$} \\
  &  (log L$_\odot$) &  (log M$_\odot$) & (km~s$^{-1}$) &  (W~Hz$^{-1}$)
  \\ \hline 
NGC\,383  & 11.66 & 11.69 & 278    & $3.4\times 10^{24}$\\    
NGC\,507  & 11.69 & 11.55 & 308    & $3.7\times 10^{22}$\\
NGC\,533  & 11.72 & 11.56 & 273    & $2.2\times 10^{22}$\\
NGC\,741  & 11.81 & 11.65 & 295    & $7.2\times 10^{23}$\\
NGC\,1407 & 11.48 & 11.37 & 273    & $6.8\times 10^{21}$\\
NGC\,2300 & 11.24 & 11.15 & 261    & $4.8\times 10^{20}$\\
HCG\,42   & 11.71 & 11.57 & 320    & \ldots \\
MKW\,4    & 11.90 & 11.77 & 275    & $1.9\times 10^{22}$\\
NGC\,4125 & 11.27 & 11.12 & 227    & $1.7\times 10^{21}$\\
NGC\,4325 & 11.35 & 11.26 & \ldots & \ldots \\
HCG\,62   & 11.62 & 11.65 & 232    & $1.6\times 10^{21}$\\
NGC\,5044 & 11.33 & 11.13 & 237    & $5.0\times 10^{21}$\\
NGC\,5846 & 11.51 & 11.39 & 237    & $2.3\times 10^{21}$\\
NGC\,6338 & 11.78 & 11.67 & \ldots & $1.1\times 10^{23}$\\
NGC\,7619 & 11.55 & 11.42 & 322    & $7.3\times 10^{21}$\\
\hline 
\end{tabular} 
\end{table}

Based on the parametrized Si/Fe ratios in Fig.~\ref{fig,norm_bin} and
the adopted SN yields, the abundance profiles have been decomposed
into sample-averaged contributions from the two SN types in
Fig.~\ref{fig,SN}.  Errors in this diagram were propagated from the
adopted statistical errors on our parametrizations of $Z_{\rm Fe}(r)$
and $Z_{\rm Si}(r)$ but are dominated at most radii by those related
to the SN yields. The latter were evaluated by varying the assumed SN
yields according to the different SN~Ia and SN~II models tabulated by
\citet{iwam99} and \citet*{gibs97}, respectively, subject to the
constraint $Z_{\rm Si}/Z_{\rm Fe} > 2.0$ for SN~II, as required by our
results at large radii in Fig.~\ref{fig,norm_bin}. At each radius, the
resulting maximum deviation relative to the case of our ``reference''
yields was taken as the uncertainty associated with the assumed SN
yields and then added in quadrature to the statistical errors on the
abundance parametrizations. The shaded regions in Fig.~\ref{fig,SN}
outline the resulting lower and upper limits. The large uncertainties
for Si in particular derive mainly from the significant variations in
predicted SN~Ia Si yields (0.14-0.27~M$_\odot$) among the models
considered. Nevertheless, at most radii $r\ga 0.1r_{500}$ considered
in this study, the resulting uncertainties can be accounted for by
assuming a characteristic error of 35~per~cent, illustrated by the
fiducial error bars in Fig.~\ref{fig,SN}. Wherever relevant, we have
taken this uncertainty into account in the error budget by adding it
in quadrature to existing errors, for example when computing total
metal masses contributed by either SN type. We also stress that while
Fig.~\ref{fig,SN} remains useful as a representation of the
sample-averaged results, the SN decomposition for individual groups
was performed using the actual measurements pertaining to each group,
with the results in the Figure being employed only when extrapolation
to $r_{500}$ was necessary.

In accordance with results for clusters, Fig.~\ref{fig,SN} establishes
on a firmer basis the results of \citet{fino00} that SN~Ia products in
groups are centrally concentrated, contributing about 80~per~cent of
the Fe in group cores, and that SN~II products on average are more
uniformly distributed, showing only mildly increased central
levels. Fe produced by SN~II therefore dominates at large radii both
in groups and clusters. Fe enrichment levels in the central metal
excesses are comparable to those in cool-core clusters and are clearly
dominated by SN~Ia as also found in clusters \citep{degr04,depl07},
but cannot be entirely explained by a central excess in SN~Ia products
alone unless assuming non-standard SN~Ia yields. This agrees with the
earlier result of \citet{fino99} based on a sample of only three
groups, as does the derived SN~II contribution to the Fe abundance at
large radii of 0.05--0.1~Z$_\odot$.

\begin{figure} 
\mbox{\hspace{-1mm} 
\includegraphics[width=83mm]{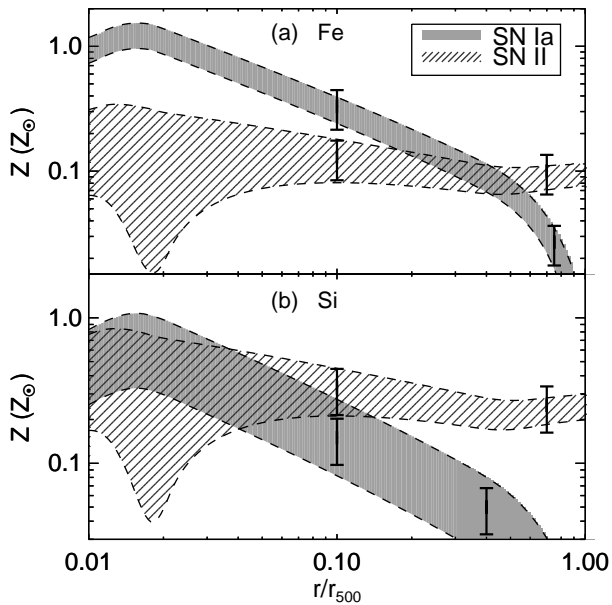}}
\caption{Sample-averaged contributions to the total abundance of (a)
   Fe and (b) Si from SN~Ia and SN~II, based on our $Z(r)$
   parametrizations. Error bars at $0.1r_{500}$ and at large radii
   show the adopted characteristic uncertainties of 35~per~cent.}
\label{fig,SN} 
\end{figure}

Total gas masses normalised by optical luminosity are summarized in
Fig.~\ref{fig,gas}. The figure reveals large variations in the ratio
of gas mass to $B$-band light, and suggests a systematic dependence on
$\langle T\rangle$, albeit one subject to significant scatter
(Kendall's $\tau=0.35$, correlation significance $\sigma_K = +1.8$,
which reduces to $\sigma_K = 1.6$ if using $L_K$ instead of
$L_B$). Inferred values range from 3--30 M$_\odot$/L$_\odot$, with a
mean of $12\pm 2$~M$_\odot$/L$_\odot$ for groups in the $T=1-2.2$~keV
range. Although the latter value is in good agreement with results of
cosmological simulations \citep{rome06}, it is nominally somewhat
lower than the corresponding value of $\approx 20$~M$_\odot$/L$_\odot$
found by \citet*{fino03}, and the overall implication of
Fig.~\ref{fig,gas} is that gas mass-to-light ratios in our groups are
well below typical cluster values.

\begin{figure} 
 \includegraphics[width=80mm]{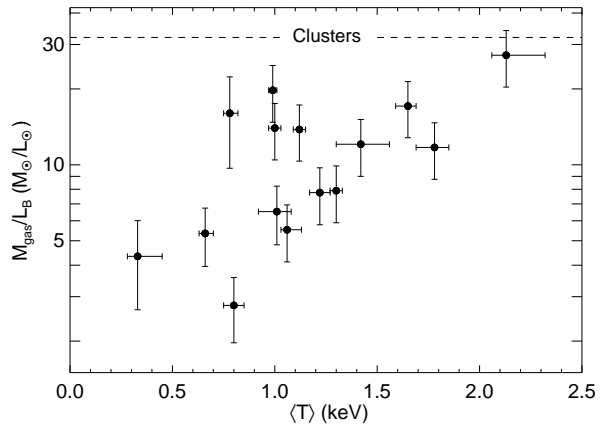}
 \caption{Gas mass-to-light ratio inside $r_{500}$. Dashed line shows
    the mean value of the $T>4$~keV clusters of \citet{fino03}.}
\label{fig,gas} 
\end{figure}

\subsection{Metal masses and mass-to-light ratios}

The results for the total X-ray metal masses in the groups are
summarized in Fig.~\ref{fig,all}, including the decomposition into
contributions from SN~Ia and SN~II. Also shown are the corresponding
metal mass-to-light ratios, all presented as a function of mean group
temperature $\langle T\rangle$.  We note that $\langle T \rangle$
correlates with total $L_B$, $L_K$, $M_\ast$, and velocity dispersion
$\sigma_{\rm v}$ at 3.6$\sigma$--3.8$\sigma$ significance, suggesting
that none of these parameters is a seriously biased proxy for total
group mass. Also note that we generally normalise by $L_B$ rather than
$L_K$ or stellar mass in these plots in order to enable
straightforward comparison to existing results in the
literature. Since total $L_K$ and $L_B$ are tightly correlated for the
sample ($\sigma_K = 4.9$), this particular choice should have little
bearing on the qualitative trends or the degree of intrinsic scatter
seen in the figures.

\begin{figure*} 
\hspace{0mm} 
\includegraphics[width=178mm]{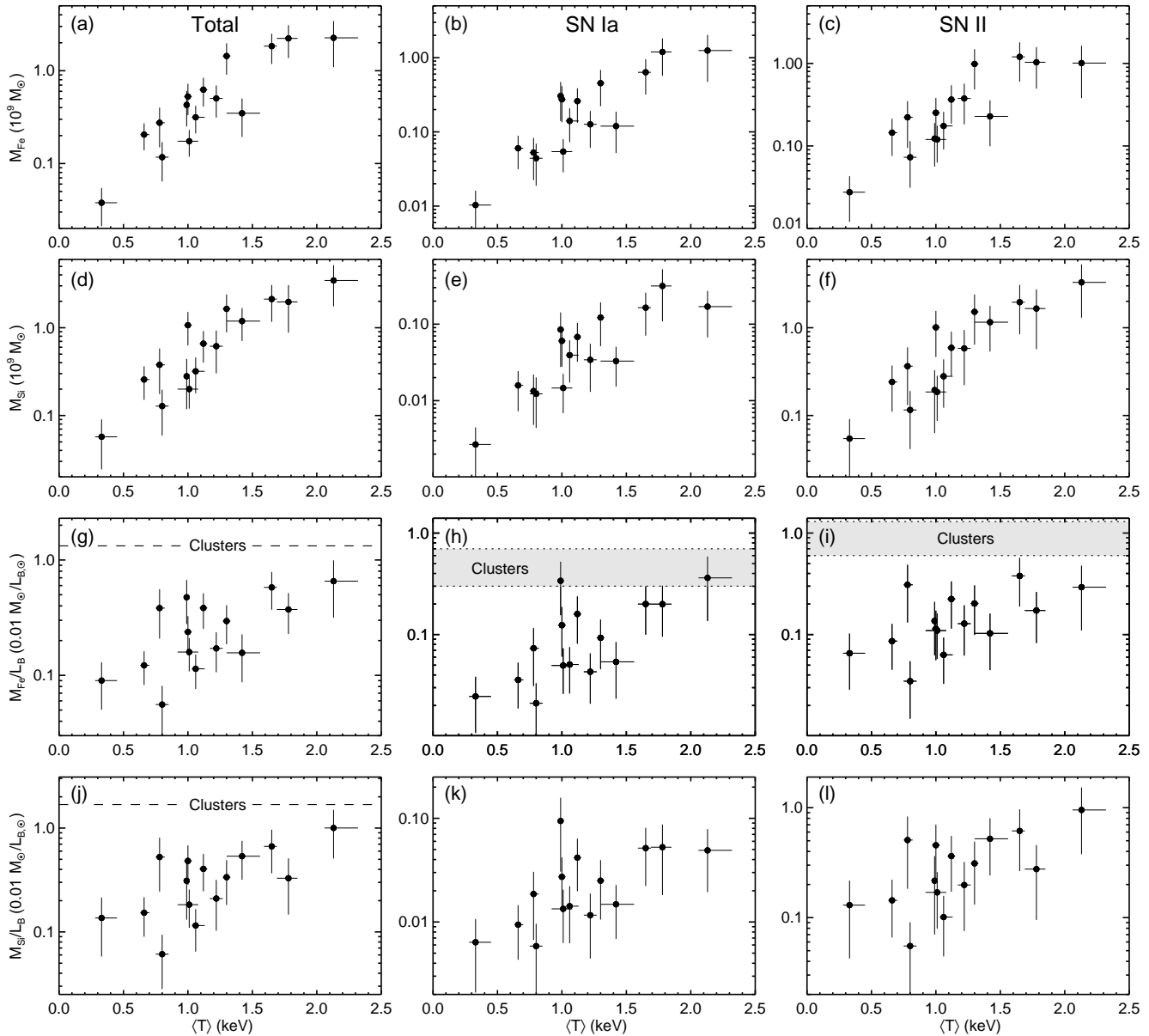}
\caption{Left panel: Derived total masses of Fe and Si inside
  $r_{500}$ as a function of mean group temperature $\langle T
  \rangle$, along with the corresponding $B$-band mass-to-light
  ratios. Centre and right panels show the results decomposed into
  contributions from SN~Ia and SN~II, respectively. Dashed lines show
  the mean values for the $T>4$~keV clusters of \citet{fino03}, and
  shaded regions mark the typical cluster ranges from \citet{fino00}.}
\label{fig,all} 
\end{figure*}

As can be seen from Table~\ref{tab,gas} and Fig.~\ref{fig,all}, the
total Fe and Si masses derived inside $r_{500}$ span two orders of
magnitude across the sample, roughly from $M \sim 3\times
10^7$--$3\times 10^9$~M$_\odot$ for both elements, and with a clear
dependence on total group `mass' $\langle T \rangle$ as anticipated.
For Si, the higher abundance observed at large radii (cf.\
Fig.~\ref{fig,norm_bin}) roughly compensates for its lower atomic mass
relative to Fe, yielding rather similar aggregate masses for the two
elements in most cases.  The fractional contribution by SN~Ia to the
metal production as listed in Table~\ref{tab,gas} implies that, on
average, SN~II have generated $\sim 60$~per~cent of the Fe and $\sim
90$~per~cent of the Si inside $r_{500}$. There is a weak tendency for
these fractions to decrease with $\langle T\rangle$, $\sigma_K = -1.6$
($\sigma_K=-2.1$ if excluding the clearly discrepant NGC\,4325),
suggesting that enrichment in cooler systems is slightly more
SN~II--dominated.

The iron mass-to-light ratio (IMLR) represents a key quantity for
studies of enrichment in groups and clusters, as it is expected to
reflect the ability of a system to generate and retain enriched
material within its gravitational potential.  As already mentioned,
the derived Fe abundances of our groups at large radii, where most of
the gas resides, are lower than the corresponding cluster values by a
factor of $\sim 2$.  As discussed in Paper~I, a similar deficit, if
slightly smaller, is also observed for Si.  The combination of a lower
metal content in the hot gas and generally lower gas mass-to-light
ratios in groups compared to more massive clusters (cf.\
Fig.~\ref{fig,gas}; see also \citealt*{lin03}) should manifest itself
in lower total metal $M/L$ ratios in cool systems.  For iron, we note
that a similar conclusion was already put forward by \citet{renz97} on
the basis of {\em ROSAT} and {\em ASCA} data collected from the
literature.  However, this study used global abundance measurements
out to the radius of X-ray detection, thus neglecting the important
fact that X-ray emission in groups is generally detected to relatively
smaller radii than clusters. Since cumulative $M_{\rm Fe}/L_B$ ratios
tend to rise outwards because the gas generally has a flatter
distribution than that of the optical light (e.g., \citealt{fino00};
\citealt{sato09}), the bias thus introduced will result in lower
values in X-ray faint systems, even if the intrinsic Fe $M/L$ ratio
inside a fixed fraction of the virial radius did not vary between
groups and clusters. The problem is only aggravated by the fact that
groups typically have shallower gas distributions than clusters and so
contain a relatively larger fraction of their hot gas (and metals) at
large radii. A fair comparison between systems of differing X-ray
brightness therefore requires all systems to be consistently compared
within a fixed overdensity radius.

These issues prompted us to investigate aggregate metal mass-to-light
ratios of our groups for comparison to clusters. The results within
$r_{500}$ are also shown in Fig.~\ref{fig,all}. The inferred ratios
vary by a factor of $\sim$~10 across the sample, with indications of
both $M_{\rm Fe}/L$ ($\sigma_K=2.1$) and $M_{\rm Si}/L$
($\sigma_K=2.3$) systematically increasing with group temperature.
Note that this is despite the fact that our temperature-independent
$Z(r)$ parametrizations, adopted for the groups for which the
abundance measurements do not extend to $r_{500}$, will tend to
suppress trends with $\langle T \rangle$ (the main motivation for
showing Fig~\ref{fig,all} in its current form is to facilitate
comparison to more massive systems, for which results are typically
provided within $r_{500}$). The results clearly suggest lower Fe (and
Si) mass-to-light ratios in cooler systems, with both ratios only
approaching typical cluster values at the high--$T$ end covered by our
sample. We note that this apparent shortfall of metals in groups is
not simply an artefact of our choice of cluster comparison data. For
example, the $T>3$~keV clusters of \citet{degr04} display an IMLR
within $r_{500}$ of $1.7\pm 0.3\times 10^{-2}$ for our adopted value
of $H_0$, even higher than the $\approx 1.3\times 10^{-2}$ reported by
\citet{fino03}.
Using a considerably larger group sample than that of \citet{fino00},
we thus confirm their tentative result that groups generally have Fe
$M/L$ of $\la 0.006$ at all radii $r\la r_{500}$ and tend to show
lower $M_{\rm Fe}/L_B$ than clusters at a fixed overdensity radius.

In order to infer the origin of the lower metal mass-to-light ratios
in cool systems, a key question to address is whether the shortfall of
Fe compared to clusters is present both for SN~Ia and SN~II products.
In Fig.~\ref{fig,all}h and i, the IMLR has been decomposed into
contributions from SN~Ia and SN~II (denoted IMLR$_{\rm Ia}$ and
IMLR$_{\rm II}$ in the following). This clearly reveals a deficit of
Fe from {\em both} SN types in groups, a conclusion also reached by
\citet{fino99}, possibly suggesting a combination of responsible
mechanisms. We note that \citet{fino99} and \citet{fino01} see a clear
increase in IMLR$_{\rm II}$ with $T$ for clusters ($T>2$ keV), and a
global deficit of SN~II ejecta in groups (as do we), but their cluster
IMLR$_{\rm Ia}$ is nearly constant with $T$ across the full range
covered by their systems, at variance with our results in the group
regime.

\subsection{Dependence on ICM temperature}\label{sec,ZT}

Figs.~\ref{fig,gas} and \ref{fig,all} suggest that the observed
variation in IMLR among the groups is driven primarily by a
systematically increasing gas mass-to-light ratio with mean group
temperature. In order to also investigate any temperature-dependence
on abundance structure, the binned abundance profiles of
Fig.~\ref{fig,norm_bin} were split into results for `cool' ($T\le
1.1$~keV) and `warm' ($T>1.1$~keV) groups in Fig.~\ref{fig,split}.
\begin{figure} 
  \includegraphics[width=80mm]{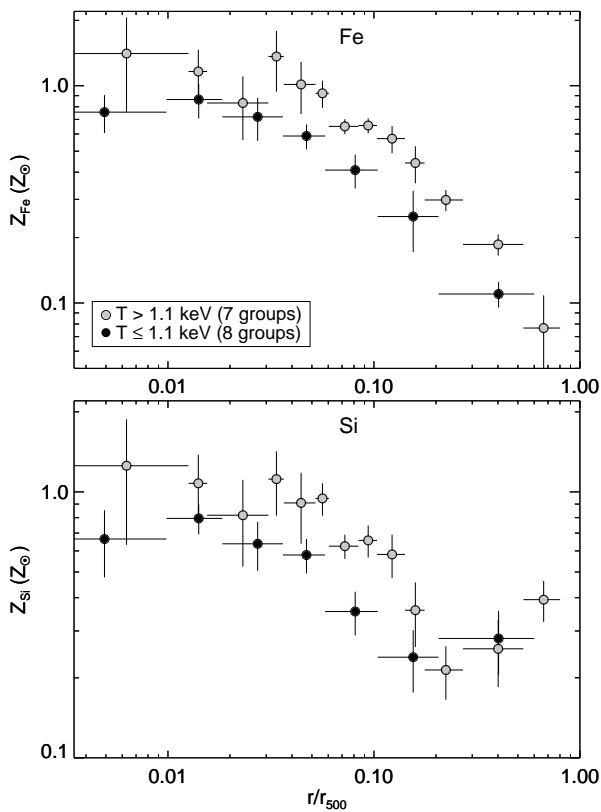}
\caption{Stacked Fe and Si profiles for `cool' ($\langle T \rangle \le
  1.1$~keV) and `warm' ($\langle T \rangle > 1.1$~keV) subsamples
  separately, in radial bins of 10 data points.}
\label{fig,split} 
\end{figure} 
We repeat here that $\langle T \rangle$ was derived within the fixed
radial range $0.1-0.3r_{500}$ and so is not affected by the strong
central cooling seen in most of the groups (cf.\ Paper~I).  Evidently,
Fe abundances in cooler groups tend to be lower at all radii common
for the two subsamples. This conclusion is clearly robust inside $r\la
0.5 r_{500}$, as will be demonstrated quantitatively below, but the
situation is less clear at larger radii where only a few of the `cool'
groups within our sample have useful {\em Chandra} data.  The results
for Si are qualitatively similar to those of Fe in the group cores ($r
\la 0.1-0.2 r_{500}$), but at larger radii there are no obvious
differences between the two subsamples. Note also that the poorer
statistics resulting from dividing the sample masks the clear
indication of a central abundance dip seen in Fig.~\ref{fig,norm_bin}.

Fig.~\ref{fig,split} indicates a general trend of higher metal
abundances in hotter groups, at least in their central regions. To
explore this quantitatively while also accounting for any systematic
differences in the amount and distribution of gas with system
temperature, we consider in Fig.~\ref{fig,ZT} the Fe mass-to-gas
ratios, effectively the mass-weighted Fe abundances.
\begin{figure} 
\hspace{-1mm}
  \includegraphics[width=85mm]{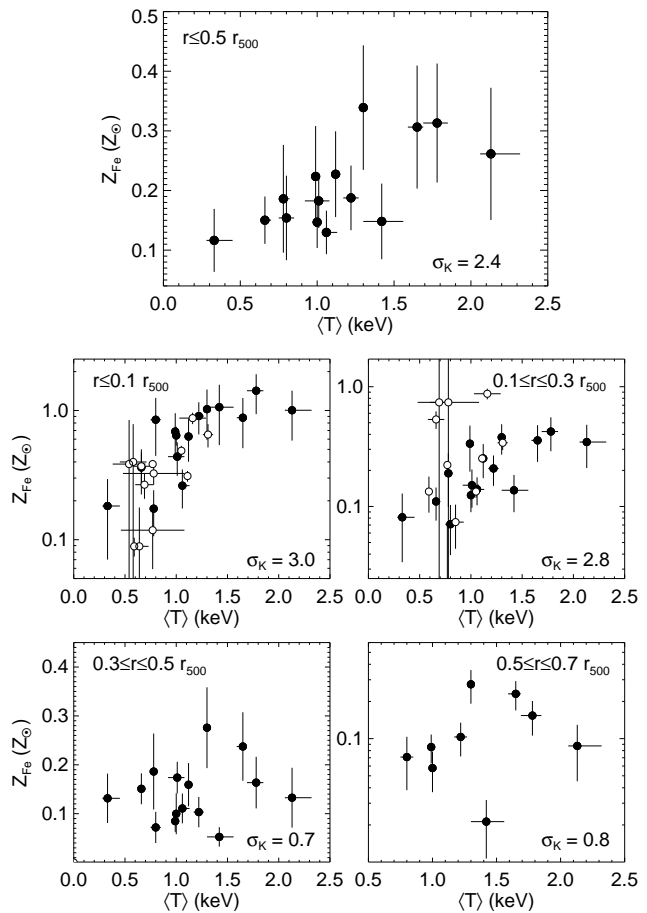}
\caption{Mass-weighted Fe abundances within various radial ranges. Our
  results are marked with filled circles, with the results of
  \citet{fino06,fino07} included as empty circles where relevant. Only
  groups with data covering the full radial interval are shown in each
  plot. Labels in lower right corners specify the significance
  $\sigma_K$ of any linear correlation among our own data points.}
\label{fig,ZT} 
\end{figure} 
The top panel shows the ratios within $0.5r_{500}$, for which we have
measurements for all groups and do not have to rely upon the
(temperature--independent) radial abundance parametrizations.  Within
this radius, there is evidence for a systematic increase in mean
$Z_{\rm Fe}$ with $\langle T\rangle$ at a moderate significance level
of 2.4-$\sigma$.  The following panels break this down into separate
radial intervals, only including groups with data in the relevant
range in each plot. For the relevant radial ranges, we have also
included the 14 {\em XMM} groups of \citet{fino06,fino07} that are not
part of our sample and for which results have been extracted in a
similar way. There is convincing evidence of an increasing Fe
abundance with system temperature in the cores of these groups, but
the trend progressively disappears as one moves to larger radii, and
is significant at less than $1\sigma$ beyond $0.3r_{500}$. This
conclusion is only strengthened by the inclusion of the
Finoguenov~et~al.\ data points in the correlation analyses, with which
$\sigma_K$ rises to $+4.1$ within $0.1r_{500}$ and drops to $+2.1$
within $0.1\le r \le 0.3r_{500}$.  The overall picture for Si is very
similar, with a clear positive correlation within $0.1r_{500}$
($\sigma_K=+2.5$) but none further out. We emphasize that the absence
of a clear trend at large radii supports our use of a single
parametrization to describe the abundance profiles beyond the radius
of X-ray detection where necessary. Outside the core, similar results
have been reported by \citet{tamu04} for $T$ and $Z$ derived within
the range 50--200~$h^{-1}$~kpc in a combined sample of groups and
clusters, and by \citet{fino01} for $Z_{\rm Fe}$ of $T\sim 1$~keV
systems derived at $0.2r_{180}$. However, at small radii the clear
trend seen here differs from results for clusters; for example
\citet{degr04} find no clear correlation between $Z$ and $T$ measured
inside $r_{2500}$ for their $T>3$ keV systems.

The observed trend in the group cores could potentially by induced by
systematics; as suggested by the work of \citet{buot00a}, a $Z$--$T$
correlation could arise due to the Fe and Si biases, which are more
pronounced at low $T$. However, for most groups, the results in the
core are based on mass-weighting spectral results extracted within
narrow radial ranges in which $T$ does not vary much. In the few cases
where it does, two-temperature models were employed in spectral
fitting, as described in Paper~I, suggesting that we have generally
resolved any temperature structure that could be affecting the
abundance determinations.  Further, the same overall trends are seen
for the emission-weighted abundances, using $Z$ directly from spectra
extracted across the relevant radial range (which in some cases
require two-temperature models for a satisfactory fit). This argues
against a systematic effect related to the modelling of the density
profiles in the group cores.  We therefore conclude that the inferred
trends should be reasonably robust against such biases.

\subsection{Supernova feedback}\label{sec,SN}

In the context of elucidating the nature and history of galactic
feedback in the groups, the SN contribution to ICM energetics is an
important quantity. Based on the adopted SN yields, we computed the
energy $E_{\rm SN}$ released by the two SN types, assuming that energy
has been released to the ICM in proportion to the metals at a level of
$10^{51}$~erg per supernova. Under this simplifying assumption, the
specific SN energy -- i.e.\ the total energy normalised by the number
of ICM particles -- released within $r_{500}$ is shown as a function
of group temperature in Fig.~\ref{fig,E_SN}. In a study of cool
clusters out to $0.4r_{180}$, \citet{fino01} found a clear increase in
specific $E_{\rm SN}$ with $T$ across the range $T=1$--3~keV, with the
trend flattening above $T\ga 3$~keV at 1.2~keV~particle$^{-1}$ (shown
as a dotted line in Fig.~\ref{fig,E_SN}). Our results sample the
low-$T$ end better and reveal no significant trend within $r_{500}$
($\sigma_K=+0.1$), although a positive trend of weak significance
($\sigma_K \approx +1$) is found within $0.6r_{500} \approx
0.4r_{180}$. Mainly as a consequence of the factor of two lower
abundances in outskirts, the sample mean of $E_{\rm SN}$ is
correspondingly lower for our groups than found for the \citet{fino01}
clusters. Note that Fig.~\ref{fig,E_SN} also suggests that $E_{\rm
SN}$ becomes comparable to the thermal energy $E_{\rm th} = (3/2)kT$
of the ICM particles below $T \sim 0.5$~keV. While this is in
excellent agreement with results of cosmological simulations
incorporating galactic outflows \citep*{dave08}, our results are
inconsistent with the specific $E_{\rm SN}$ declining with $\langle T
\rangle$ as predicted by those simulations.

\begin{figure} 
\mbox{\hspace{0mm} 
  \includegraphics[width=84mm]{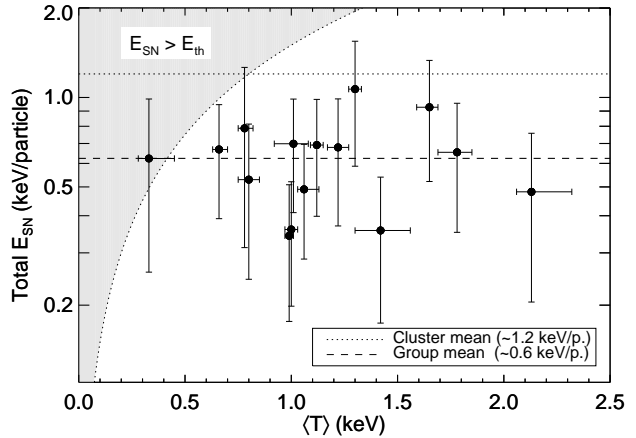}} 
\caption{Released SN energy per ICM particle inside $r_{500}$. Dashed
  line marks our mean value of $E\approx 0.6$~keV~particle$^{-1}$,
  while dotted line shows the mean found for $T\ga 3$~keV clusters
  \citep{fino01}. Shaded area shows the region where $E_{\rm SN}$
  exceeds the thermal energy $E_{\rm th}$ of the ICM.}
\label{fig,E_SN} 
\end{figure} 

The total SN energy released in the groups spans the range $4\times
10^{59}$--$2\times 10^{61}$~erg.  Although SN~Ia contribute about
$\sim 20$~per~cent of the $E_{\rm SN}$ within the group core, the vast
majority ($\sim 95$~per~cent) of the total energy released within
$r_{500}$ has been provided by SN~II. We emphasize however, that these
numbers, as well as those shown in Fig.~\ref{fig,E_SN}, simply
represent the SN explosion energy {\em associated} with the metals in
the ICM and not necessarily the energy {\em delivered} by SN to the
hot gas. The two quantities would differ if, for example, the fraction
of synthesized metals reaching the ICM is different from the
corresponding value for the SN explosion energy. We will return to
these issues in Section~\ref{sec,release}, noting for now that
uncertainties on the inferred $E_{\rm SN}$ are not necessarily
dominated by those associated with our gas mass estimates or the
choice of SN model yields.

\subsection{The origin of the central iron excess}\label{sec,central}

The presence of a central Fe excess in these groups parallels results
for clusters which show this feature to be common in cool-core systems
(e.g., \citealt{fino00}; \citealt{degr04}). As for clusters, the
excess can extend well beyond the optical extent of the central galaxy
and can show a dip at the very centre, inside $r\la 0.5D_{25}$, where
$D_{25}$ is the ellipse outlining a $B$-band isophote of 25
mag~arcsec$^{-2}$. As shown in Fig.~\ref{fig,SN}, the enrichment in
group cores has not been exclusively provided by SN~Ia, but the Fe
{\em excess} itself, relative to the Fe level at large radii, has
largely been provided by SN~Ia. This is broadly consistent with the
idea that the central early-type galaxy is responsible for the Fe
excess. While evidence for this conclusion is also observed in
clusters (e.g.\ \citealt{boeh04,degr04}), this scenario has not been
tested in any detail in groups. One could {\em a priori} expect a
similar situation to hold for lower-mass systems, since the fractional
contribution of the brightest galaxy to the total light is generally
larger in these than in clusters \citep{lin04}.

In order to investigate this, we computed the mass of Fe produced by
SN~Ia within $0.1r_{500}$. On average, this corresponds to a radius of
$r\approx 50$~kpc, roughly matches the radial extent of the cool core
where present (cf.\ Paper~I), and corresponds to twice the maximal
$D_{25}$ extent of the central galaxy and to the radius where the
radially weakly declining SN~II contribution to Fe production becomes
consistent with the level in the outskirts (Fig.~\ref{fig,SN}).
Results were normalised by the $K$-band luminosity of the central
galaxy from Table~\ref{tab,BGG}. Assuming our adopted SN~Ia yields
along with the observed present-day SN~Ia rate in nearby early-type
(E/S0) galaxies of $R_0 = 3.5^{+1.3}_{-1.1}\times
10^{-14}$~yr$^{-1}$~L$_{K,\odot}^{-1}$ \citep{mann05}, the resulting
time-scales required for SN~Ia in the central galaxy to generate the
observed Fe mass are illustrated in Fig.~\ref{fig,MFe_Lk}a. Results
from Si broadly agree with those from Fe. As can be seen, there is a
clear dependence of the Fe $M/L$ ratio, and hence required enrichment
time-scale, on $\langle T \rangle$ within $0.1r_{500}$, qualitatively
mirroring the global behaviour inside $r_{500}$ (cf.\
Fig.~\ref{fig,all}). Both the mean and median enrichment time-scales
for the sample are $\sim 5$~Gyr, which, given the increasing
time-scale with $T$, is entirely consistent with the corresponding
results of \citet{boeh04}, who find that times of $\ga 5$~Gyr are
required in clusters.

\begin{figure} 
 \hspace{0mm} 
 \includegraphics[width=80mm]{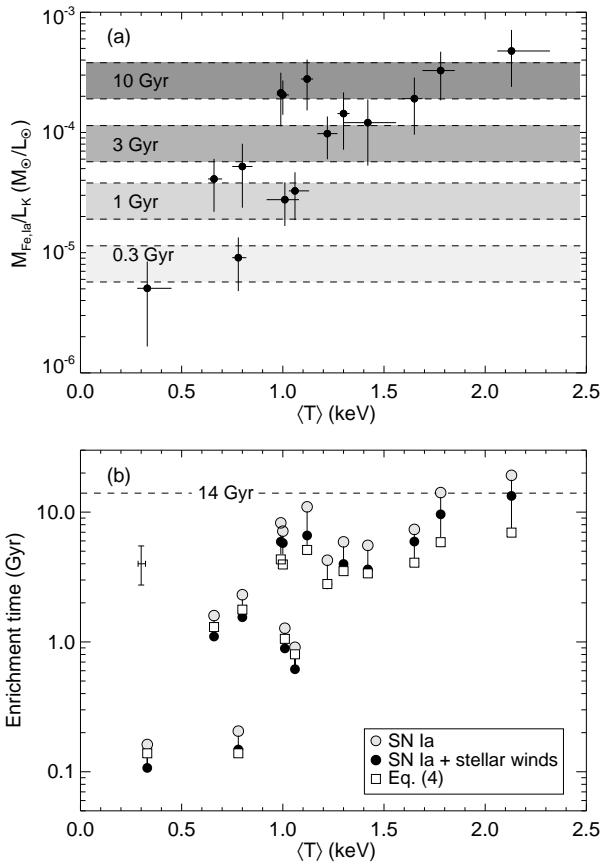}
 \caption{(a) Mass of Fe produced by SN~Ia within $0.1r_{500}$,
   normalised to the $K$-band luminosity of the central group
   galaxy. Shaded regions outline the errors on the typical
   time-scales required for SN~Ia to produce the observed $M_{\rm
   Fe}/L_K$ ratios, assuming a constant SN~Ia rate with redshift.  (b)
   The corresponding enrichment times, including also the effect of
   stellar mass loss in the BGG, and that of assuming an SN~Ia rate
   evolving with redshift according to equation~(\ref{eq,snia}). Lines
   connect the time-scales for each group as computed under these
   different assumptions. A typical error bar (dominated by
   uncertainties in the local SN~Ia rate) is shown at $T=0.3$~keV.
   Dashed line marks an enrichment time-scale of 14~Gyr.}
\label{fig,MFe_Lk} 
\end{figure}

The immediate implication is that in many of these groups, SN~Ia in
the central group galaxy could well be responsible for the Fe excess
seen within the cool core. Some caveats to this interpretation,
however, include the facts that our method does not include the Fe
that has become locked in stars over the relevant time-scales
(although this might be a small fraction if star formation largely
ceased at high redshift in these central galaxies), nor do we account
for any Fe expelled beyond $0.1 r_{500}$ by galactic winds, or for the
growth of stellar mass ($\sim L_K$) of the BGG over the considerable
time-scales involved. The inferred time-scales may thus be viewed as
lower limits under the adopted assumptions. On the other hand, there
could be a contribution to central enrichment due to stellar mass loss
in the BGG and due to gas lost to group cores from satellites via
galactic winds or environmental interactions. A more sophisticated
approach would thus include the contribution from stellar winds along
with a model of the stellar mass evolution of the BGG. An argument for
not considering the latter effect here, however, is the observation
that the bulk of the stellar mass in many central cluster ellipticals
was in place already at $z=2-3$ (e.g., \citealt{stot08}), so this may
not play an important role. A case specific to our sample is
NGC\,1407, in which the BGG appears to have formed over half its
stellar mass already by $z\approx 5$ \citep{spol08}.

Mass loss due to stellar winds can be trivially incorporated in our
calculations, and it is instructive to consider the impact of
this. For this purpose, we consider {\em all} the Fe present within
$0.1 r_{500}$ (i.e.\ not just that produced by SN~Ia), but subtract a
uniform base level of 0.1~Z$_\odot$ as seen in the group outskirts to
obtain an estimate of the {\em excess} iron mass within this radius.
Following \citet{boeh04}, the stellar mass loss rate $\dot M_\ast$
from the central early-type is then estimated from
\begin{equation}\label{eq,winds}
  \frac{\dot M_\ast}{\mbox{ M$_\odot$~yr$^{-1}$}} \approx 1.5\times
  10^{-11} \frac{L_B}{\mbox{L$_\odot$}}
  \left(\frac{t_\ast}{\mbox{15~Gyr}}\right)^{-1.3} ,
\end{equation}
where $t_\ast$ is the `age' of the galaxy \citep{ciot91}. The rate of
Fe loss is obtained by multiplying this estimate by the mean Fe mass
fraction in stellar winds, which in turn depends on stellar
metallicity.  Assuming mildly supersolar metallicities for the BGG
stellar population ($Z=1.5$~Z$_\odot$, corresponding to [Z/H]~$\approx
+0.2$; see e.g.\ \citealt{sanc07}) along with our adopted Solar
abundance table \citep{grev98} and a typical galaxy age of 10~Gyr
\citep{sanc06}, the prediction of equation~(\ref{eq,winds}) is added
to the SN~Ia iron production rate and the associated enrichment times
plotted in Fig.~\ref{fig,MFe_Lk}b.

On average, the inferred stellar mass loss can account for
$\sim$~25~per~cent of the Fe within $0.1r_{500}$, reducing the
required enrichment times correspondingly. Nominally, this brings the
estimated time-scales below 14~Gyr for all groups, alleviating -- but
not obviating -- the need for additional sources of metal enrichment
in the cores of high--$T$ groups. Hence, subject to the adopted
assumptions, the time-scales may still be prohibitively long for
systems with high $M_{\rm Fe}/L_K$ ratios in the core. Tentative
evidence that satellite galaxies may have contributed to building up
the central metal excesses in at least one of our groups is provided
by \citet{mend09}, who demonstrate a good correspondence between the
radial X-ray Fe distribution and that of stellar mass density in the
NGC\,5044 group out to $r\approx 100$~kpc, corresponding to five times
the $D_{25}$ extent of the BGG. Yet, barring a significant additional
contribution to central Fe enrichment from other group galaxies, we
are compelled to conclude that our approach generally underestimates
the metal production of the BGG over cosmic time.  Possible loopholes
include invoking an SN~Ia rate per unit $L_K$ that evolves with
redshift, similar to what has been proposed for early-types in more
massive clusters (e.g., \citealt{renz93,boeh04}), or a significant
contribution to central ICM enrichment either from intracluster stars
(see Sec.~\ref{sec,release}) or from the sinking of highly enriched
low--entropy gas towards group cores \citep{cora06}.

We can briefly explore the former possibility, the effect of a
redshift-dependent SN~Ia rate. To this end, we assume an evolving
SN~Ia rate $r_{\rm Ia}(t)$ per rest-frame time $t$ and co-moving
volume as described by the star formation rate SFR$(t)$ at a given
time convolved with the SN~Ia delay-time distribution $\Phi(t)$,
\begin{equation}
  r_{\rm Ia}(t) \propto \int_{t_0}^{t} \mbox{SFR}(t')\Phi(t-t')\mbox{d}t',
\label{eq,snia}
\end{equation}
where $t_0$ is the age of the Universe at the initial redshift of star
formation, here assumed to be at $z_f = 10$.  Following
\citet{etto05}, SFR$(t)$ is adopted from the extinction-corrected
model of \citet{stro04}, and $\Phi$ is assumed to be a Gaussian of
characteristic delay time $\tau=4$~Gyr and $\sigma = 0.2\tau$ (the
results are not highly sensitive to the exact choice of $\Phi$; see
\citealt{etto05}). In this model, the SN~Ia rate peaks at $z\approx
0.8$ at $\approx 4.8$ times the present-day value. Excluding here the
effects of stellar winds (for straightforward comparison to the
results assuming a constant SN~Ia rate with no stellar wind
contribution), the resulting enrichment time-scales are plotted as
empty squares in Fig.~\ref{fig,MFe_Lk}b.  This modification is seen to
reduce the enrichment time-scales to $\sim 5$--7~Gyr even for the
hottest groups, in agreement with similar estimates for cluster BCGs
\citep{boeh04}.  Thus, in this model, the central Fe excess can easily
have been generated by the BGG itself within half a Hubble time,
without the need for additional sources of central enrichment.

A final point worth mentioning here, related to mass loss through
stellar winds, involves the associated abundance ratios. Since neither
Si nor Fe is processed in the intermediate-mass stars primarily
responsible for such winds, the $\alpha$/Fe ratio of these winds (and
of their parent stars) will be identical to that of the gas from which
the stars formed. This ratio can generally be expected to be high in
early-type galaxies, where star formation was largely terminated at
high redshift as indicated by the supersolar $\alpha$/Fe~ratios
typically found for the stellar population in these galaxies (e.g.,
\citealt{sanc07}).  Consequently, stellar winds from our BGGs should
have $\alpha$/Fe ratios approaching those of SN~II ejecta.
Interestingly, Fig.~\ref{fig,SN} indicates a mild central enhancement
in SN~II abundances which is at least consistent with a significant
contribution from stellar mass loss associated with the BGG. Whether
this feature can be entirely explained by stellar mass loss is a
complicated question, however. A contribution from SN~II to central
enrichment is certainly still required, as the predicted sample mean
of $Z_{\rm Si}/Z_{\rm Fe} \approx 0.95$ from SN~Ia and stellar winds
alone falls short of the observed mass-weighted mean of $Z_{\rm
Si}/Z_{\rm Fe} \approx 1.10$ inside $0.1r_{500}$. The radial extent of
the SN~II enhancement suggests either a contribution from stars not
currently in the BGG, i.e.\ other group members or intracluster stars,
or the action of some gas and metal mixing within $\sim 0.3
r_{500}$. Moreover, the spatial distribution of stellar mass loss
products is not straightforward to predict, since both [Z/H] and
[$\alpha$/Fe] in early-type stellar populations tend to show radial
gradients; while [Z/H] generally peaks towards the galactic centre,
$\alpha$/Fe gradients may be positive, negative, or flat, and the mean
absolute value may depend on galaxy mass \citep{sanc07,spol09},
implying that the Fe fraction and abundance ratio in stellar winds may
be a highly non-trivial function of radius and of the environment of
the BGG. A detailed exploration of this would likely require X-ray
data of a superior quality and is beyond the scope of this study.

\section{Origin and fate of ICM metals in groups}\label{sec,fate}

We can briefly summarize the results of the previous Section as
follows. Cool-core groups tend to show similar levels of central
enrichment to those of more massive systems, but abundance gradients
outside the core are stronger, leading to lower abundances at large
radii. Group outskirts are heavily dominated by SN~II enrichment, with
the total energy per ICM baryon injected by SN somewhat lower than in
more massive systems. At the same time, the global gas mass for a
given amount of stellar light is lower in cooler systems within
$r_{500}$, leading to lower global metal mass-to-light ratios, both
for SN~Ia and SN~II products. The latter result in particular may hold
important information on the enrichment history of these systems. In
this Section we investigate a range of possible explanations for this
while acknowledging that this is not necessarily an exhaustive list.
Fundamentally, these explanations can be divided into mechanisms
causing (i) less efficient metal generation or metal transfer to the
ICM in low-mass systems, (ii) removal of metals from groups over
cosmic time, or (iii) cosmological accretion of less enriched gas in
small systems.

\subsection{Galaxy formation efficiency and star formation history}
\label{sec,SFE}

As an example of a mechanism belonging to the first category mentioned
above, we first consider the possibility of a reduced galaxy formation
efficiency at low $\langle T \rangle$.  Everything else being equal,
this would lower the amount of Fe released per unit ICM mass, and
although this cannot in itself explain the observed IMLR variations
(unless the stellar initial mass function also varies with
environment), it could help explain the globally lower metal
abundances seen in cooler systems.

We can test this scenario on the basis of the gas mass-to-light ratio
in Fig.~\ref{fig,gas}, or equivalently, the ratio of stellar mass
$M_\ast$ to that of baryons, $M_{\rm bar}\approx M_\ast + M_{\rm ICM}$
(sometimes denoted the cold fraction $f_c$, or the star formation
efficiency) shown in Fig.~\ref{fig,starbar}. This plot indicates a
general decrease in cold fraction with $\langle T\rangle$
($\sigma_K=-1.5$, strengthening to $\sigma_K=-2.2$ if excluding the
somewhat peculiar NGC\,2300; see Paper~I), but with a tendency for
this quantity to be higher than corresponding results for clusters
\citep{etto03}. Hence, the galaxy formation efficiency actually
appears to {\em rise} in lower-mass systems within our sample, in
concordance with other observational and theoretical findings
(\citealt{lin03}; \citealt{rome06}; \citealt*{gonz07};
\citealt{dave08}).
\begin{figure} 
\mbox{\hspace{-1mm} 
 \includegraphics[width=84mm]{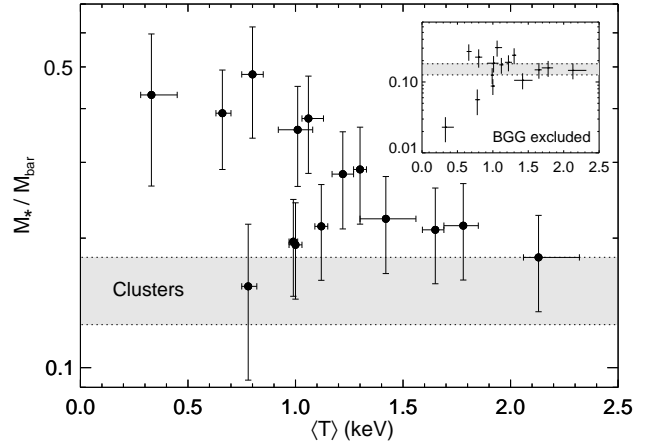}}
 \caption{Stellar-to-baryon mass ratio inside $r_{500}$. The shaded
   region outlines the 1-$\sigma$ range of $M_\ast /(M_\ast + M_{\rm
   ICM}) = 0.16^{+0.02}_{-0.03}$ covered by the clusters of
   \citet{etto03}. Inset shows the corresponding result (same units)
   with the contribution from the BGG excluded.}
\label{fig,starbar} 
\end{figure} 
Hence, there is no relative dearth of stellar mass within our groups,
and we can readily exclude more exotic scenarios to account for the
lower IMLR such as the possibility that the stellar population has
simply produced systematically fewer metals in cool groups; the
metallicity of the stellar population is typically at least $\sim$
solar in many of the bright early-types in our groups, with no obvious
differences from central cluster galaxies (see, e.g.,
\citealt{hump06,brou07}). However, it is also clear that the moderate
trend of decreasing cold fraction with $T$ is largely driven by the
BGG (see inset in Fig.~\ref{fig,starbar}). If considering the
contribution to $M_\ast /M_{\rm bar}$ from satellite galaxies alone,
there is no systematic variation with group temperature
($|\sigma_K|=0.3$, regardless of whether NGC\,2300 is included), and
most groups are then consistent with the corresponding cluster
results.

For a fixed baryon fraction, a higher cold fraction in low-mass
systems is perhaps no surprise. Cooler systems have lower
characteristic entropies \citep*{ponm03} and hence cooling times
\citep*{sand06}, and so a larger fraction of their hot gas can be
expected to cool out. As discussed by \citet{gonz07}, a higher
galaxy/star formation efficiency in low-mass systems could also be
promoted by an increased efficiency of tidal interactions (boosting
star formation rates) and a decreased efficiency of strangulation,
i.e\ the stripping of any gas halo of late-type galaxies.

It is nevertheless interesting to note that, if accounting also for
the iron locked in stars, a {\em higher} $f_c$ in cool systems could
potentially maintain a {\em total} IMLR comparable to that seen in
clusters.  However, it seems that only the stellar population of the
BGG itself could likely provide any significant modifications to a
trend in this total IMLR with $T$, because Fig.~\ref{fig,starbar}
shows that the stellar metal contribution from satellite galaxies
would not vary systematically with $T$ unless stellar metallicity
does. For the sake of argument, we note that any correlation of total
IMLR with $\langle T\rangle$ can be made to vanish for our sample
provided that the stellar population in the BGG has mildly supersolar
metallicity, [Fe/H]~$\approx +0.2$ in all groups. While not an
unreasonable assumption (see, e.g., \citealt{hump06}), the resulting
mean total IMLR of $\sim 7\times 10^{-3}$ would still be a factor of
$\sim 3$ below the corresponding value for clusters (which is $\sim
0.02$ for our adopted abundance table; \citealt{renz04}). The
contribution from satellite galaxies can raise this mean, but restores
the temperature dependence, unless the (mass-weighted) mean stellar
metallicity declines from $Z_{\rm Fe} \sim 2$~Z$_\odot$ to
$\sim$\,solar across our temperature range.  In other words, a higher
$f_c$ in cooler groups can only keep the {\em total} IMLR independent
of $T$ and in accordance with cluster values while maintaining the
trends in Fig.~\ref{fig,all} if a much larger fraction of synthesized
metals remains locked in stars in cool systems. Although we lack the
data to exclude this with confidence, it seems unlikely for several
reasons, in part because it requires considerable systematic variation
in star formation histories and/or IMF and stellar binary fraction
among the bright early-types in cool and hot groups. Secondly, a
higher average stellar metallicity in cooler groups -- in which the
average galaxy stellar mass is lower, as discussed below -- would
conflict with the well-established local mass--metallicity relation,
which appears fairly insensitive to environment (e.g.,
\citealt{bern03}).

Including the contribution of any intracluster stars in our groups
could help explain the trends in Fig.~\ref{fig,starbar}, if such a
component is more prominent in poorer systems as suggested by
\citet{gonz07}. However, given that its estimated contribution to the
stellar mass in clusters is still of order $\sim 30$~per~cent (see
Section~\ref{sec,release}), it seems improbable that it can account
for the order-of-magnitude variation in $M_{\rm gas}/L_B$ among our
groups, nor would it explain the seriously reduced IMLRs in cool
systems. In summary, it seems clear that a systematically varying
galaxy formation efficiency, or even a contribution from stars not
included in our estimates of $L_B$ or $M_\ast$, cannot account for the
results in Fig.~\ref{fig,all}. Given the higher cold fraction in
cooler systems, one would naively expect higher ICM abundances in
these, at variance with our results.

Although a varying star formation efficiency cannot explain the trends
in Fig.~\ref{fig,all}, it is still instructive to also explore the
potential role of variations in average star formation {\em history}
among the groups. For example, is it conceivable that ICM enrichment
has somehow been sufficiently `delayed' in lower-mass systems to help
explain their lower IMLR? To investigate this, we show in
Fig.~\ref{fig,SiFe} the mass-weighted Si/Fe ratio of the groups as a
function of the corresponding Fe abundance, both restricted to radii
$r\le 0.5 r_{500}$ inside which we have reliable measurements for all
15 groups. Systems have been grouped according to their mean
temperature. Nominally, the data show a negative correlation with
$Z_{\rm Fe}$ at the 2.0-$\sigma$ level, suggesting a higher Si/Fe and
lower overall Fe abundance in cooler groups within these radii. The
statistical errors are large, however, clearly exceeding the
systematic variations, so the results have also been accumulated in
three temperature bins for clarity, as illustrated in the figure
inset. This confirms the initial impression, demonstrating that the
ICM in cooler and less enriched groups tends to contain a relative
preponderance of SN~II ejecta.
\begin{figure} 
\mbox{\hspace{0mm} 
 \includegraphics[width=82mm]{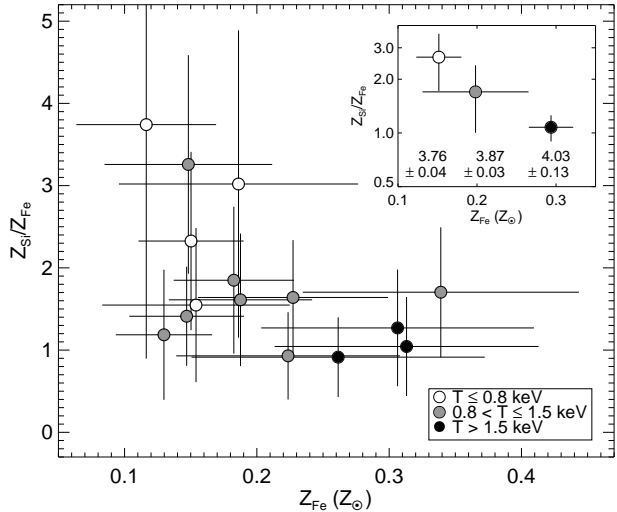}}
 \caption{Mass-weighted Fe abundances and Si/Fe ratios within
  $0.5r_{500}$.  Systems are colour-coded according to their $\langle
  T\rangle$ as labelled. Inset shows mean results within the
  temperature bins, with uncertainties reflecting the mean and
  standard deviation within each bin. Annotations in inset provide the
  mean total $B$--$K$ colour within $r_{500}$ and its 1-$\sigma$ error
  for the group members within each bin.}
\label{fig,SiFe} 
\end{figure} 

In analogy with stellar population studies, the tendency for the ICM
to be less Fe-enriched and display higher $\alpha$/Fe ratios in cooler
groups suggests a chemically `younger' ICM in these groups. On
average, galaxies also tend to be slightly bluer in the cooler groups,
as indicated in the figure inset, and furthermore are less massive,
with the average stellar mass from equation~(\ref{eq,mstar}) showing a
positive correlation with $\langle T\rangle$ at 3.0 and 3.4-$\sigma$
significance for BGGs and satellites, respectively. While several
explanations may be possible for these results, the data are at least
consistent with a scenario in which the (generally bluer and less
massive) galaxies in our cooler groups have experienced slightly more
extended star formation histories, whereas the conversion of the
available gas into stars was completed earlier in the more massive
galaxies in hotter groups. As a consequence, star formation and metal
production (from SN~Ia in particular) might be progressively `delayed'
in lower-mass systems.

Although these results should be regarded as tentative ($\sigma_K =
-1.3$ for a correlation between $Z_{\rm Si}/Z_{\rm Fe}$ and total
$B$--$K$ colour), they could be a manifestation of the observation
that more massive early-type galaxies tend to be older, i.e.\ have
completed the bulk of their star formation (and the associated ICM
enrichment by SN~II and stellar winds) at higher redshift
(`downsizing', e.g., \citealt{catt08}), as well as an effect by which
such galaxies at fixed mass tend to be slightly older in more massive
environments (e.g.\ \citealt{sanc06}), perhaps indicative of an
environmental dependence of downsizing. This interpretation is also in
accord with the inference of \citet{loew06} that the fraction of
baryons processed through stars is lower in lower-density environments
compared to in rich clusters. Nevertheless, the inferred average
colour variations between galaxies in the cool and hot groups are
small, $\Delta(B-K) \la 0.3$, so it remains highly questionable
whether average star formation histories have varied sufficiently with
group mass to explain the observed order-of-magnitude variation in
IMLR from both SN~Ia and SN~II.

In conclusion, it seems unlikely that arguments invoking changing star
formation efficiencies {\em or} histories with group mass can
themselves explain the results in Fig.~\ref{fig,all}.  The above
results are, on the other hand, consistent with, for example,
substantial gas (and hence metal) loss from within $r_{500}$ in
lower-mass systems. In fact, the simulations of \citet{dave08}
indicate that the higher cold fraction in cool systems (which,
naively, could be expected to result in higher ICM metal abundances),
is compensated for by galactic outflows that drive more metals out of
the potentials of cooler systems. We consider this possibility in
Section~\ref{sec,gasloss}.

\subsection{Metal release efficiency}\label{sec,release}

Metals can be released into the ICM through several mechanisms,
including galactic outflows, stripping of enriched material by
galaxy--galaxy and galaxy--ICM interactions, and direct {\em in situ}
enrichment by intracluster stars. Here we consider these mechanisms in
turn, posing the question of whether metals have been released less
efficiently into the ICM in cooler groups. One immediately attractive
feature of this possibility is its potential to account for the lower
metal mass-to-light ratios of both SN~Ia and SN~II products.  For
example, gaseous stripping would likely affect both SN~Ia and SN~II
products, and tidally stripped intracluster stars should release both
SN~Ia and SN~II metals with equal efficiency. In addition, any metal
release mechanism whose efficiency scales with system temperature
might alleviate the need for a redshift-dependent SN~Ia rate in the
BGG (Section~\ref{sec,central}), and help explain the observed
$Z$--$T$ trend in group cores.

\subsubsection{Galactic outflows}

Some fraction of all metals released by SN in the group members will
never mix with the ICM but will remain in the parent galaxy and
potentially participate in subsequent star formation.
Fig.~\ref{fig,all} then raises the possibility that SN ejecta were
released less efficiently into the ICM in cooler systems. To evaluate
the feasibility of this scenario, we can compare the total SN energy
inferred from the ICM abundance pattern $E_{\rm SN,ICM}$
(Fig.~\ref{fig,E_SN}) to that expected from the stellar population
within a group, $E_{\rm SN, \ast}$.  Naively, one would expect
$\eta=E_{\rm SN,ICM}/E_{\rm SN, \ast} \la 1$, the exact value
depending on the mass fraction of SN ejecta retained by the galaxies.

For each group, $E_{\rm SN, \ast}$ can be estimated from the stellar
mass of the group members, noting that a Salpeter IMF with an assumed
minimum mass for core-collapse progenitors of 8~M$_\odot$ implies one
core-collapse SN per $\sim 150$~M$_\odot$ of stars formed. In
addition, a typical stellar $\alpha$/Fe ratio of $\sim$~solar in the
group members \citep{hump06} implies that at least $\sim 0.3$~SN~Ia
must have exploded for each SN~II. Adding these numbers yields $E_{\rm
SN, \ast} \sim 9\times 10^{48} (M_\ast$/M$_\odot)$~erg , and the
resulting comparison to $E_{\rm SN,ICM}$ within $r_{500}$ is plotted
in Fig.~\ref{fig,release}. With a resulting sample mean and 1-$\sigma$
error of $\eta= 0.68\pm 0.10$, our numbers are in broad agreement with
existing theoretical and observational estimates of the mass loss
fraction of galaxies, which range from $\eta \approx 0.4-0.8$ (see,
e.g., \citealt{siva09} and references therein). In particular, as
illustrated in Fig.~\ref{fig,release}, the latter authors find a value
of $\eta = 0.84^{+0.11}_{-0.14}$ within $r_{500}$ in clusters using an
approach based on the observed SN~Ia rate in cluster early-types,
while noting that the fraction could be as low as 35~per~cent if the
assumed SN rates were increased to within their 2-$\sigma$ upper
limit.

\begin{figure} 
 \hspace{0mm} \includegraphics[width=82mm]{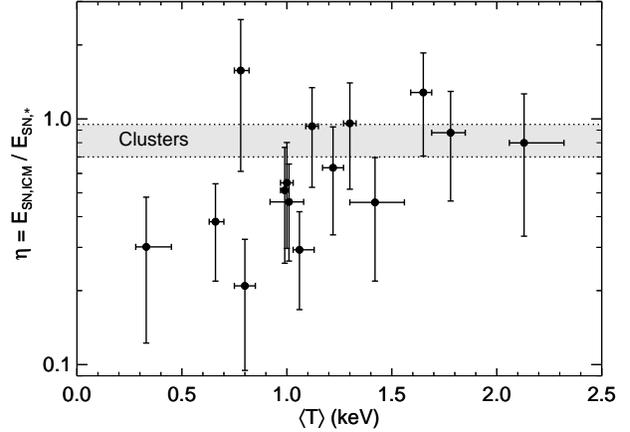}
 \caption{Ratio of SN energy released in the groups as inferred from
    ICM abundances to that expected from the stellar mass of the group
    members. Shaded region marks the estimate for the $T\approx
    2$--6~keV clusters of \citet{siva09}.}
\label{fig,release} 
\end{figure}

Interestingly, while this suggests that ICM enrichment has been fairly
efficient for our sample as a whole, there {\em is} also a tendency
($\sigma_K=+1.7$, rising to $+2.5$ if excluding the outlying
NGC\,2300) towards higher $\eta$ in hotter groups. This result could
be anticipated from Figs.~\ref{fig,E_SN} and \ref{fig,starbar}: Since
the estimated $E_{\rm SN,ICM}$ per ICM particle does not vary
systematically with $T$, the trend in Fig.~\ref{fig,release} is a
consequence of the systematic decrease in cold fraction with $T$. The
result is consistent with the notion that SN products have been
released more efficiently into the ICM in hotter systems, although
other explanations are possible. These include the possibility of the
stellar IMF and/or mean metallicity varying systematically with $T$,
as this would lead to systematic variations in the predicted number of
SN per stellar mass and in the estimated stellar mass itself.

Nominal values of $\eta$ exceeding unity are seen in a few cases, but
all results are consistent with $\eta < 1$ even without accounting for
uncertainties in SN rates and yields. Other possible effects include
the presence of optically undetected intracluster stars, which would
contribute to $E_{\rm SN,ICM}$ only, and more speculatively the
possibility of pre-enrichment, whereby some of the ICM material
accreted by the groups was enriched by galaxies not currently in the
group. In addition, we cannot exclude that ICM masses may be slightly
overestimated in some groups, nor that the SN~Ia contribution to
$E_{\rm SN, \ast}$ may be underestimated (cf.\ \citealt{siva09}). The
fact that $\eta$ is not constant and generally {\em different} from
unity is no surprise though. As pointed out by \citet{dave08}, X-ray
emission-weighted $\alpha$/Fe ratios, used here to infer $E_{\rm
SN,ICM}$, may not be a good indicator of the relative number of SN~Ia
and SN~II that have exploded within the group, but rather reflects the
metal distribution mechanisms at work.

A possible interpretation of the systematic trend in
Fig.~\ref{fig,release} is that SN--driven outflows have been more
efficient at releasing metals into the ICM in hotter groups. Whether
this might be expected on the basis of the higher average galaxy mass
in such groups is not immediately obvious though. A deeper galactic
gravitational potential might disfavour the development of a freely
outflowing starburst wind, but, conversely, star formation may
generally have proceeded more rapidly and vigorously in massive
galaxies. An alternative explanation invokes the possibility of
AGN--driven outflows helping to release metals into the ICM.
Observational evidence for such outflows to exhibit significantly
supersolar abundances for a range of elements has been reported
\citep{arav07,wang07}, and numerical simulations \citep{moll07}
indicate that AGN outflows could contribute significantly to cluster
enrichment.

Since most of our groups show some evidence of ongoing or recent AGN
activity (Paper~I), the relevant question is whether such activity,
and its impact on the ICM, scales with $\langle T\rangle$ as required
by Fig.~\ref{fig,release}. More specifically, since AGN activity may
expect to scale with stellar bulge mass of the host galaxy, we can
test whether the trend in Fig.~\ref{fig,release} is explicable on the
basis of higher BGG bulge masses in hotter groups. Adopting again the
distinction between cool and hot groups introduced with
Fig.~\ref{fig,split}, we note that $\eta$ is significantly higher in
hotter ($T>1.1$~keV) groups by $58 \pm 18$~per~cent. Using the BGG
central stellar velocity dispersions $\sigma_{0,\ast}$ listed in
Table~\ref{tab,BGG} we find some tendency for these to increase
systematically with $T$ at the same level of significance as that of
$\eta$ vs.\ $T$ ($\sigma_K=+1.7$), However the trend is considerably
flatter than that of $\eta(T)$, with results showing a mean and
1-$\sigma$ error of $\langle \sigma_{0,\ast} \rangle = 267\pm 15$ and
$278 \pm 10$~km~s$^{-1}$ for cool and hot groups, respectively,
indistinguishable at the 1-$\sigma$ level. Qualitatively similar
results are obtained for $\langle \sigma_{0,\ast} ^4 \rangle$, which
is expected to scale with the mass of the central supermassive black
hole \citep{ferr00,gebh00}. In other words, there are no indications
that central black hole masses are generally higher in hotter groups,
and this result still ignores the fact that relatively more gas needs
to be enriched in more massive systems. Hence, unless more massive
black holes are considerably more efficient {\em for their mass} at
expelling metals from their host bulges, these considerations suggest
that the trend in Fig.~\ref{fig,release} is not driven by increased
AGN activity in hotter groups, at least not if associated with the BGG
only.

\subsubsection{Stripping and interactions}

Gas and metal loss from galaxies induced by interactions with the ICM
could also help explain the trend in Fig.~\ref{fig,release}. An
indication that some (presumably enriched) gas has been removed from
our group galaxies by other means than SN outflows is provided by the
factor of 2--3 deficiency in H{\sc i} mass derived for the spirals in
NGC\,5044 and NGC\,7619 compared to field spirals of similar optical
morphology and size \citep{seng06}. In addition, evidence for ongoing
or past stripping associated with galaxy--ICM interactions is present
in a few of our groups \citep*{rasm06,mend09}. General support for
this interpretation comes from the work of \citet{fino06} who find a
larger scatter in metallicity than in entropy in the group cores
within their sample, which they interpret as evidence for diversity in
the origin of gas in the cores. On the other hand, \citet{renz04}
notes that if, e.g., ram pressure stripping were important for ICM
enrichment, then hotter clusters with higher galaxy velocity
dispersion should show relatively higher ICM abundances. While this is
not observed in his results, nor in those of \citet{baum05}, such a
trend {\em is} seen for our sample, and, interestingly, is confined to
the regions of highest ICM density (Fig.~\ref{fig,ZT}).

This trend is certainly consistent with a density--dependent metal
release mechanism in group cores, whose efficiency scales with galaxy
velocity dispersion. If ram pressure stripping is the culprit, we
might expect the ICM abundance in the central region to scale more
strongly with the characteristic ram pressure in this region than with
group temperature. In order to test this possibility, the velocity
dispersion $\sigma_{\rm v}$ of each group (Table~\ref{tab,gas}), was
used to estimate the characteristic ram pressure within
$0.1r_{500}$. Assuming $P_{\rm ram} = (M_{\rm gas}/V) \sigma_{\rm
v}^2$, with $V=(4/3)\pi (0.1r_{500})^3$, we find evidence for a
moderately significant correlation between $P_{\rm ram}$ and the
mass-weighted Fe abundance within $0.1r_{500}$ ($\sigma_K = +2.1$),
though the correlation of $Z_{\rm Fe}$ with $\langle T\rangle$ is
actually stronger [$\sigma_K = +2.6$ for own groups, $+3.0$ when
including those of \citet{fino06,fino07} as discussed in
Section~\ref{sec,ZT}].  In addition, the corresponding correlation of
$Z_{\rm Si}$ with $P_{\rm ram}$ is weak ($\sigma_K=+1.1$ compared to
$+2.5$ for $Z_{\rm Si}$ vs.\ $\langle T\rangle$). While keeping in
mind that the estimated $\sigma_v$ may not always be representative of
the typical 3-D galaxy velocity in each group if affected by velocity
anisotropies or small-number statistics (although 13 out of our 15
groups have more than 10 members), we thus find that results are
consistent with ram pressure stripping contributing to enrichment in
group core, but the evidence is tentative at best, and we cannot rule
out other density-dependent mechanisms as more important. Moreover, as
this process should be inefficient at large radii (and where, indeed,
no clear trend in $Z$ with $T$ exists), we conclude that the reduced
efficiency of stripping through galaxy--ICM interactions in cooler
systems is unlikely to account for their lower IMLRs or for the trend
in Fig.~\ref{fig,release}.

Note that if the metallicity--temperature trend in group cores does
reflects an increased efficiency of environment--driven gas loss from
galaxies in hotter groups, then the absence of a similar trend in
systems with $T\ga 3$~keV \citep{degr04} may suggest that such
processes `saturate' at $T\sim 2$--3~keV, perhaps as a consequence of
near--complete gas removal from core galaxies in clusters above these
temperatures. It has yet to be established to what extent this
possibility is in accord with simulation results, which suggest that
only a minor fraction ($\sim 10$~per~cent) of the global ICM metal
budget even in massive clusters can be accounted for by ram-pressure
stripping, though with the fraction considerably higher in cluster
cores \citep{doma06, kapf07}.

Other interaction processes involve galaxy--galaxy interactions such
as harassment, tidal interactions, and mergers. The latter process in
particular may also release metals to the ICM, in part because a
significant fraction of a galaxy's ISM may shock-heat and become
unbound during major mergers, and in part because the merger activity
may further trigger a starburst-- or AGN--induced outflow (e.g.,
\citealt{cox06,hopk06}). Our data do not allow useful constraints on
the contribution of this process to ICM enrichment, however, and
cosmological simulations may still lack the dynamic range to robustly
assess whether it has been substantially less efficient or influential
in low-mass systems compared to clusters. Naively, one would expect
tidal stripping and mergers to be more effective in poorer systems
owing to the lower velocity dispersion, suggesting that these
mechanisms cannot explain the trends in Figs.~\ref{fig,all} and
\ref{fig,release}.

\subsubsection{Enrichment by intracluster stars}

Intracluster (IC) stars represent a non-negligible component of the
baryons in groups and clusters, and could pollute the ICM very
efficiently since their ejecta are released directly into the
ICM. Observations suggest that $\sim 30$~per~cent of the stellar mass
in clusters within $r_{500}$ may be contained in IC stars
\citep{gonz07}, but it remains debatable whether this fraction
decreases with total cluster mass as claimed by \citet{gonz07}, or
rises asymptotically with mass to a level of $\sim 50$~per~cent of
$M_{\ast}$ as suggested by other observational and theoretical studies
\citep*{lin04,conr07,purc07}. 

Since this component is not accounted for in the cluster comparison
results in Figs.~\ref{fig,gas} and \ref{fig,all}, including its
contribution to $L_B$ could affect the apparent group--cluster
discrepancies. For example, if IC light is less prominent in cooler
systems as suggested by some of the above results, this could help
explain the lower ICM abundances in groups. However, even a
50~per~cent increase in IC stellar light in massive clusters is
insufficient to bring the lower group IMLR within the factor of two of
rich clusters that is explicable on the basis of the lower
large--radius abundances in groups. Furthermore, \citet{siva09}
conclude that 20--25~per~cent of the Fe in the ICM within $r_{500}$
can be accounted for by IC stars, implying that IC stars contribute in
roughly equal proportions to the stellar mass and ICM Fe enrichment in
clusters. This suggests that the resulting IMLR should not be
significantly affected by the presence of this
component. Consequently, it seems we can exclude the possibility that
intracluster light provides an important contribution to the trends in
Fig.~\ref{fig,all}. This would be particularly true if this component
is actually more important in cooler systems, as suggested by
\citet{gonz07}.

IC stars could still play an important role for ICM enrichment in
group cores, however.  A relevant question is whether the metal
contribution from these can alleviate the need for a
redshift-dependent SN~Ia rate in the BGG to account for the Fe within
the cool core (cf.\ the discussion of Fig.~\ref{fig,MFe_Lk}). Where IC
light in clusters can be separated from that of the central galaxy,
the former typically contributes $\sim 80$~per~cent of their combined
optical output within $r_{500}$ \citep{siva09}. This fraction will be
lower inside $0.1r_{500}$, because IC light is more widely distributed
than that of the BGG. For simplicity, if assuming that the IC light
traces the group gravitational potential and that its $K$-band
luminosity density is distributed according to an NFW profile
\citep*{nava97} with a typical concentration $c= 5$--10, then
5--10~per~cent of this light would fall inside $0.1 r_{500}$. The
implication is that this component can potentially reduce the
enrichment times in Fig.~\ref{fig,MFe_Lk} by a factor $\sim 1.4$
(optimistically assuming a total $K$-band contribution of four times
that of the BGG, of which 10~per~cent falls within
$0.1r_{500}$). Although this factor assumes that none of this light is
already accounted for in the adopted values of $L_{K,\rm BGG}$, it may
still be large enough to eliminate the need for an evolving BGG SN~Ia
rate, especially if the IC contribution does increase with group
mass. What IC stars probably {\em cannot} do, however, is account for
the observed $Z$--$T$ trend in group cores (Fig.~\ref{fig,ZT}). Even
if the fractional Fe contribution from IC stars within $0.1r_{500}$
increases by 50~per~cent across our temperature range (at variance
with the \citealt{gonz07} results), this would still fall an order of
magnitude short of explaining the factor $\sim 10$ rise in $Z_{\rm
Fe}$ within this radius. Arguments invoking a far superior enrichment
efficiency from this component (compared to stars in galaxies) would
also fail, since Fig.~\ref{fig,release} suggests that galaxies are
already fairly efficient in polluting the ICM.

\subsection{Cooling of enriched gas in group cores}

Turning now to the possibility that metals were {\em removed} from
groups rather than not produced or released into the ICM, we first
consider a scenario in which radiative cooling has removed enriched
gas from the X-ray phase. At a given fraction of $r_{500}$, cooling
times tend to be lower in cooler systems (e.g., \citealt{sand06}), so
this explanation would be consistent with such systems exhibiting
lower $M_{\rm gas}/L_B$ and missing relatively more metals, as
suggested by Figs.~\ref{fig,gas} and \ref{fig,all}. The process is
nevertheless still limited by the observation of the central abundance
excess, because cooling rates above a certain threshold would
eventually destroy the central excess due to the associated inflow of
less enriched material from larger radii.

Radiative cooling would primarily deplete SN~Ia products, due to their
higher concentration at small radii where cooling is most rapid.
Thus, the two questions we seek to address is whether the amount of Fe
generated by SN~Ia that is `missing' in Fig.~\ref{fig,all}h (or its
$K$-band analogue) can be accounted for by cooling, and whether that
would ultimately destroy the central Fe excess in conflict with our
observations. Since Figs.~\ref{fig,all} and \ref{fig,starbar} indicate
that the properties of our most massive group, the $T\approx 2.1$~keV
system NGC\,6338, are fairly representative of those of more massive
clusters, we can use this group as a benchmark for characterizing
deviations in cool systems from cluster results.  On the basis of the
$K$-band IMLR$_{\rm Ia}$ in NGC\,6338 deduced from
Table~\ref{tab,gas}, we evaluated the missing Fe mass in each group
and hence the corresponding total missing gas mass for a given assumed
Fe abundance of the cooled-out gas. Making the assumptions that the
gas cooling out had $Z\approx $~Z$_\odot$ as observed in the group
cores, and, conservatively, that this gas has been cooling for 10~Gyr
(corresponding to an initial redshift of $z=2$ for the adopted
cosmology), resulting mean cooling rates could be computed.

Fig.~\ref{fig,cc}a shows the required cooling rates as a function of
mean group temperature, demonstrating that modest average rates of a
few tens of M$_\odot$~yr$^{-1}$ would be necessary under the above
assumptions.  Note that NGC\,6338 itself has not been included in the
plot, nor has the $\langle T\rangle \approx 1$~keV system NGC\,4325,
which actually exhibits a slightly higher $K$-band IMLR$_{\rm Ia}$
than NGC\,6338 and so is not Fe--deficient according to our
assumptions.  Of course, the derived rates assume that gas in group
cores was already enriched to solar levels by $z\approx 2$ and has
been subject to unimpeded cooling ever since. These estimates are thus
likely representing a best-case scenario, and the total missing gas
mass becomes correspondingly larger if assuming a lower $Z_{\rm Fe}$
for the removed gas. For example, adopting instead the current $M_{\rm
Fe}/M_{\rm gas}$ ratio within $r_{500}$ in each group as
representative of the assumed Fe content of the cooled-out gas, the
required $\dot M$ rises to a sample mean of $\sim
130$~M$_\odot$~yr$^{-1}$. Such values seem prohibitively large at
present, but we cannot {\em a priori} exclude cooling rates of this
magnitude in the past. Note also that cooling should be more efficient
in lower-$T$ systems, and although Fig.~\ref{fig,cc}a indicates a
weakly significant tendency ($\sigma_K = +1.2$) towards higher cooling
rate requirements in hotter systems, there is a clear negative trend
with $T$ if normalising the required $\dot M$ by the total gas mass
($\sigma_K=-2.6$) or gravitating mass ($\sigma_K=-2.1$, assuming
$M_{\rm grav}\propto T^{3/2}$) of the group.

\begin{figure} 
 \mbox{\hspace{-2mm}
 \includegraphics[width=86mm]{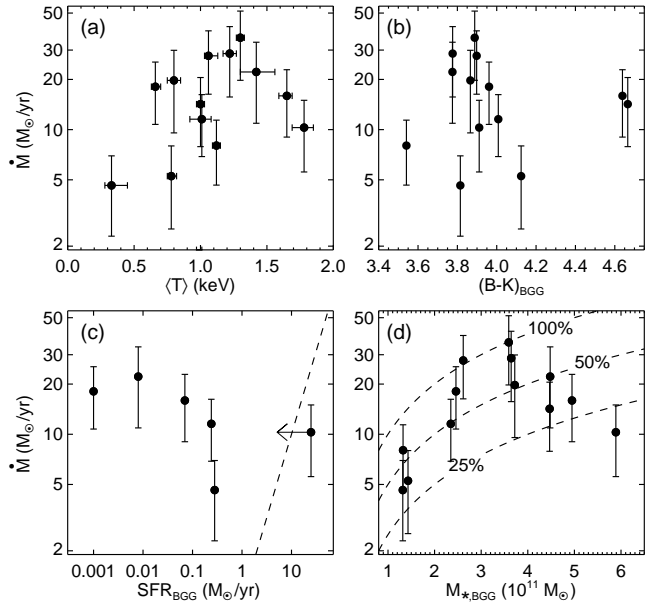}}
 \caption{The time-averaged gas cooling rate $\dot M$ required to
  account for the shortfall of IMLR$_{\rm Ia}$ compared to NGC\,6338,
  assuming cooling over a 10~Gyr time-scale. Results are plotted as a
  function of (a) $\langle T\rangle$, (b) the $B$-$K$ colour of the
  brightest group galaxy, (c) the star formation rate of the latter,
  where available (see text for details), and (d) the stellar mass of
  the latter. Dashed line in (c) represents equality, while lines in
  (d) mark out constant fractions of cooled-out gas mass relative to
  the stellar mass of the BGG, with fractions as labelled.
  Uncertainties on $\dot M$ reflect the relative errors on IMLR$_{\rm
  Ia}$.}
\label{fig,cc} 
\end{figure}

Although the central temperature is not observed to drop below $\sim
40$~per~cent of the peak value in any of the groups (Paper~I), it is
still relevant to ask whether there are any indirect signs of {\em
ongoing} strong cooling where this is most needed. If central cooling
is currently proceeding at the required rates, and some of the
cooled-out gas participates in star formation, one might expect
systematically bluer BGGs in higher--$\dot M$ groups. We investigate
this in Fig.~\ref{fig,cc}b, which reveals no significant dependence
between $\dot M$ and BGG colour ($\sigma_K=-0.7$), not even if
normalising required $\dot M$ by $L_K$ of the BGG ($\sigma_K=-0.5$).
For the galaxies with the relevant available data, we can be more
specific and compare the required $\dot M$ to the current central star
formation rate (SFR), noting that some low-level `residual' star
formation (SF) seems to be taking place in many nearby early-type
galaxies (e.g., \citealt*{comb07}). Hence, unless cooling and SF is
highly episodic, one might again expect a correlation between $\dot M$
and current SFR. We have used available H$\alpha$ and UV data from the
literature to test this scenario (again excluding NGC\,4325 and
6338). H$\alpha$-based SFRs from SDSS--DR4 data \citep{brin04} were
adopted for NGC\,4073 (the BGG in MKW\,4) and NGC\,5846, while
H$\alpha$ fluxes were taken from NED for NGC\,383 and 741
\citep*{cape05}, and UV fluxes from NED for NGC\,1407 \citep*{rifa95}
and NGC\,4125 \citep{dale07}. For the latter four galaxies, SFRs were
estimated using the relations of \citet{kenn98}, which should be
sufficiently accurate for our purposes. The UV luminosities have been
corrected for Galactic extinction but not internal dust extinction,
which should be small for these early-types. The H$\alpha$ fluxes
typically derive from the central galaxy regions, so some AGN
contribution is possible in these cases. For example, the H$\alpha$
emission from NGC\,4073 cannot be unambiguously classified as either
due to star formation or an AGN, so we have treated this result as an
upper limit. The results are plotted in Fig.~\ref{fig,cc}c and show,
if anything, a trend that runs opposite to expectations in our cooling
scenario, and with SFRs generally several orders of magnitude below
the required cooling rates.

Thus, we find no convincing evidence that the required central cooling
in the groups is associated with any recent SF in the BGG, with the
necessary cooling rates generally higher than even current SFRs in
central galaxies of `cooling flow' clusters (these rarely exceed
5--10~M$_\odot$~yr$^{-1}$ and are often much smaller; cf.\
\citealt{raff06}). This is corroborated by the fact that there is no
clear tendency for the BGG to be bluer in cooler systems ($\sigma_K =
-0.4$) where cooling is expected to be more efficient.  Further, large
amounts of recent cooling-induced star formation seems incompatible
with the inferred ages of the stellar population in many of the BGGs,
which are of order 10~Gyr (e.g., \citealt{hump06}).  This issue can
potentially be circumvented in the --~perhaps somewhat unlikely~--
event that much of the cooled-out gas has been deposited without
participating in star formation. If so there should be substantial
(several $10^{10}$~M$_\odot$) reservoirs of cold gas residing in the
group cores, but H{\sc i} measurements listed in NED, where available,
only reveal upper limits, with NGC\,5846 as the only exception.

The above considerations argue against significant recent cooling, and
reveal no indications that ongoing cooling is more important where
required. However, this does not preclude the possibility of stronger
cooling activity in the past, where it could have contributed to
building the current stellar mass of the BGG. If so, and if cooling is
indeed responsible for removing metals, one might expect the stellar
mass $M_{\rm \ast, BGG}$ of the central galaxy, and the ratio of this
to total ICM mass within $r_{500}$, to correlate positively with the
required amount of cooled-out gas. This is not clearly the case
though, with correlation significances of only $\sigma_K = +1.3$ and
$+0.1$, respectively.  An example is demonstrated in
Fig.~\ref{fig,cc}d, which compares $\dot M$ to $M_{\rm \ast, BGG}$.
In a few cases, the required cooled gas mass nominally represents
$\sim 100$~per~cent of the current stellar mass of the BGG (implying
that the vast majority of the metals proposed to have cooled out
cannot have been supplied by the BGG itself), while on average it
corresponds to about half this value.

Although the latter case may not necessarily be completely at variance
with the typical history of mass build-up and star formation in these
galaxies (which may still have assembled a significant fraction of
their current stellar mass since $z \sim 2$), there is an additional
and very powerful constraint on the allowing cooling rates arising
from the presence of the central Fe excesses. As the typical estimated
iron injection rate from the BGG is of order $\sim
0.02$~M$_\odot$~yr$^{-1}$ if including both SN~Ia and stellar winds,
the cooling rates implied by Fig.~\ref{fig,cc} would destroy the
observed excesses on typical time-scales of only $\sim 10^7$~yr
through the associated inflow of less enriched material from larger
radii (while also flattening the central entropy distribution for the
same reasons, although a discussion of this is beyond the scope of the
present study). This severely compromises the viability of recent
cooling as a mechanism to reduce the global X-ray iron mass-to-light
ratios in cooler systems.

Could central cooling and inflow of low-$Z$ gas still play a role for
establishing the central $Z$--$T$ trend in Fig.~\ref{fig,ZT}? If
repeating the cooling calculations underlying Fig.~\ref{fig,cc} using
the `missing' metal mass inside $0.1 r_{500}$ (again assuming that
IMLR$_{\rm Ia}$ in NGC\,6338 is representative of the case of no
missing metals), then the typical required cooling rates are a factor
of $\sim 3$ smaller, with some tendency for the required $\dot M /
M_{\rm gas}$ to decrease with $T$ ($\sigma_K=-2.5$) and for $\dot
M/L_{\rm K,BGG}$ ($\sigma_K=-1.7$) to decrease with ($B$--$K$)$_{\rm
BGG}$. This suggests that central cooling could have played some role
in removing metals from group cores and suppress the central Fe excess
in the coolest systems, although, again, unimpeded cooling at the
rates required to explain the $Z$--$T$ trend within $0.1r_{500}$ in
Fig.~\ref{fig,ZT} would generally have destroyed the observed Fe
excesses.

In summary, although cooling rates under optimistic assumptions {\em
may} be sufficiently low in some systems for this mechanism to remain
a viable means of removing some metals from the X-ray phase, the rates
must generally be far below those required to account for the global
shortfall of Fe within $r_{500}$ implied by Fig.~\ref{fig,all}. In
addition, there are no clear indications that cooling is, or has been,
more prominent in the systems with the largest deficit of SN~Ia
metals.  As mentioned, cooling would also primarily affect SN~Ia
products, thus likely still requiring a separate mechanism to also
account for the lower IMLR$_{\rm II}$ in cooler systems. We thus
conclude that central cooling cannot generally have been important in
establishing the low IMLR of groups.

\subsection{Recent gas and metal loss from groups}\label{sec,gasloss}

Another way to induce a lower IMLR in groups is through gas or metal
loss from the potential wells of low-mass systems. This could help
account for the lower gas mass-to-light ratios (via bulk gas loss) and
potentially also the lower abundances (via preferential loss of highly
enriched gas). As a means of discriminating between bulk gas loss and
selective loss of enriched gas, we note that only the latter process
would generally reduce the inferred abundances. The efficiency of such
a mechanism should depend on group temperature, and our sample is
large enough to allow a crude test for such an effect as demonstrated
in Figs.~\ref{fig,split} and \ref{fig,ZT}.

Beyond $0.5r_{500}$, Fe has mainly been produced by SN~II
(Fig.~\ref{fig,SN}), so $Z_{\rm Fe}$ should here provide a crude
measure of the ability of the systems to retain the products of
starburst-driven galactic winds. There is little evidence from
Fig.~\ref{fig,ZT} that this ability varies systematically across the
temperature range probed here ($\sigma_K < 1$), although the
correlation significance rises to $\sigma_K=+1.2$ if excluding the
fossil group NGC\,741 which is extremely iron-deficient in the
outskirts. Hence, although statistics at large radii are limited and
dominated by the hotter groups in the sample, with only three systems
from the 'cool' ($T\le 1.1$~keV) subsample included in the last panel
of Fig.~\ref{fig,ZT}, it seems clear that selective removal of
enriched gas from cooler systems does not contribute significantly to
the trends in Fig.~\ref{fig,all}. Moreover, given the strong radial
decline of the mean Fe profile for the sample, a possible alternative
-- {\em bulk} gas loss from outskirts without preferential loss of
metals -- would predominantly have removed low--$Z$ gas, thus leaving
material with a {\em higher} abundance behind. Hence, while such a
mechanism could have lowered $M_{\rm gas}/L_B$ in cool systems, it
cannot account for the seriously reduced IMLRs in these environments.

Gas and metals can certainly still have been lost from groups. The
absence of a clear temperature--abundance trend within
0.5--1.0~$r_{500}$ could potentially be an artefact of small--number
statistics and does not preclude the possibility that enriched gas has
been expelled well beyond $r_{500}$ in all systems.  Any such activity
might be ascribed to galactic outflows driven by powerful SN or AGN
activity, displacing ICM metals to radii where they would generally
escape detection in these low-$T$ systems. In the absence of
significant mixing, highly enriched high-entropy material released by
such outflows would rise buoyantly in the ICM until in entropy
equilibrium with the surroundings. We here consider the feasibility of
such a scenario for both SN-- and AGN--driven outflows.

\subsubsection{Supernova-driven winds}

Supernova-driven galactic winds should generally be dominated by SN~II
products. Indeed, observations of extraplanar X-ray emission
associated with hot gas around star-forming galaxies indicate
significantly enhanced $\alpha$/Fe ratios of 2--3 times the Solar
value \citep{stri04}. The radial increase in the Si/Fe ratio seen for
our groups outside the central galaxy is, in fact, only reproduced in
simulations where the feedback scheme favours the ejection of SN~II
products from galaxies over those of SN~Ia (see
\citealt{torn04,rome06}). This raises the interesting possibility that
the radial increase in the Si/Fe ratio inferred for many of our groups
may be explained by SN~II--dominated material that has risen in the
ICM to mix with it at radii corresponding to a significant fraction of
$r_{500}$, and occasionally beyond. This rise could have occurred by
means of buoyancy forces and/or bulk kinetic energy imparted to the
gas.  Since star-forming spirals tend to avoid group cores
\citep{hels03}, these galaxies will typically already be surrounded by
high--entropy gas, so convective displacement of the enriched ejecta
to large radii may be inefficient and mechanical work energetically
favoured for removing large amounts of gas from within $r_{500}$.

Stringent tests of this possibility are impeded by the fact that SN
wind properties remain poorly established. In particular, detailed
hydrodynamical simulations indicate that much of the wind fluid in an
SN--driven outflow may escape the galaxy in a high--entropy phase that
would be difficult to detect with current X-ray instrumentation
\citep{stri00}. To obtain a crude estimate of wind properties, we may
consider the tail gas being lost from the NGC\,2276 starburst galaxy,
a member of the NGC\,2300 group in our sample \citep{rasm06}. This
material is likely to contain a significant contribution from
starburst outflows, though possibly with some hot ISM and surrounding
ICM mixed in. The entropy of this gas is $\sim 300$~keV~cm$^2$,
although this figure is probably uncertain by a factor of $\sim 2$ due
to the unknown geometry of the tail along the line of sight. Taken at
face value, this gas would rise due to convection to at least $r \sim
0.6r_{500}$ if assuming the adopted ICM density profile for the group.
This consideration suggests that starburst wind material {\em can}
rise to significant radii in the ICM by means of buoyancy, provided
that mixing with the surrounding material is suppressed at least
initially, e.g., by magnetic fields.

If SN~II--enriched material from starburst winds has been
preferentially lost from lower-mass systems to generate some of the
trends in Fig.~\ref{fig,all}, one would naively expect a positive
correlation between $M_{\rm Si}/M_{\rm Fe}$ and $\langle T \rangle$.
While Fig.~\ref{fig,SiFe} already suggests otherwise, we examine this
possibility more directly in Fig.~\ref{fig,Fe_Si_T}, again showing
this ratio within $0.5 r_{500}$ only, to avoid reliance on our
temperature--independent abundance parametrizations.  The results are
compared to the analogous values derived by \citet{sato07} for the
$T\approx $~3--3.5~keV clusters AWM~7 and A1060 inside 0.35 and
$0.25r_{180}$ ($\approx 0.6$ and $0.4r_{500}$),
respectively. Fig.~\ref{fig,Fe_Si_T} indicates that Si has {\em not}
been preferentially lost from the cores of lower-mass systems for the
$T$--range probed here. To the contrary, there is mild evidence for a
negative correlation ($\sigma_K=-1.8$) although this is driven by the
three coolest systems, without which any correlation vanishes
($\sigma_K=-0.3$). There is also no significant trend at larger radii
($\sigma_K=+0.5$), as evidenced by the inset showing the corresponding
results within 0.5--0.7~$r_{500}$ for the nine groups with data
extending across this range. Everything else being equal,
buoyancy--driven bubbles of SN~II material should be able to rise to
larger radii in cooler (lower-entropy) systems, but there is no
evidence that the $M_{\rm Si}/M_{\rm Fe}$ ratio is higher at low
$\langle T\rangle$ within this radial range. In other words, there is
no clear evidence for a loss of SN~II products from the central
regions of cooler groups, nor for any such material to have been
displaced to larger radii in those systems.

\begin{figure} 
\mbox{\hspace{-2mm} 
\includegraphics[width=85mm]{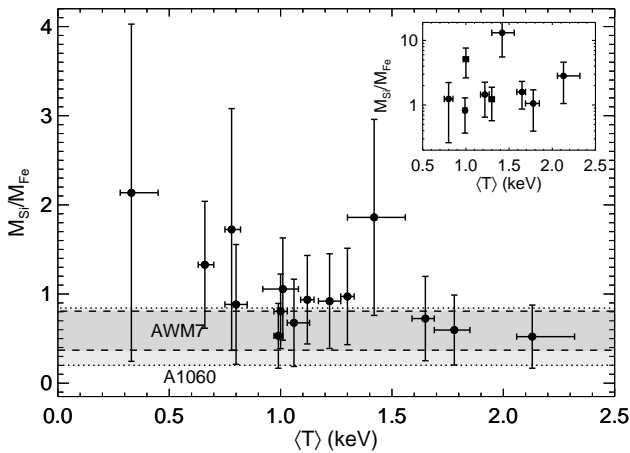}}
\caption{Ratio of silicon-to-iron mass inside $0.5 r_{500}$. Also
  shown are the 90~per~cent confidence intervals derived for the
  clusters AWM\,7 (darker shade, dashed lines) and A1060 (lighter
  shade, dotted lines) by \citet{sato07}. Inset shows the
  corresponding values within the range 0.5--0.7~$r_{500}$ for the
  groups with data across this range.}
\label{fig,Fe_Si_T} 
\end{figure}

Comparison to the \cite{sato07} results thus suggests that any
selective removal of SN~II products from the central regions of
low-mass systems by starburst winds must have occurred with comparable
efficiency in systems with $T\la 3.5$~keV, and so are unlikely to
explain the trends seen for SN~II products in Fig.~\ref{fig,all}. This
conclusion is consistent with the results seen for the combined
group--cluster sample of \citet{fino00}, which admittedly contains
only four $T<2$~keV systems.  These authors find a roughly constant
$Z_{\rm Si}/Z_{\rm Fe}$ ratio at $r=0.2r_{180}$ for systems with a
corresponding Si $M/L$ ratio or gas mass fraction $f_{\rm gas}$ below
a certain threshold. Above this value, the Si/Fe ratio appears to rise
dramatically, with the transition occurring around Si $M/L \approx
0.01$~M$_\odot$/L$_\odot$ or $f_{\rm gas}\approx 0.1$, corresponding
to $T \approx$3--4~keV \citep{fino01}. One interpretation is again
that metal loss due to starburst winds could be efficient in groups
but that this process must have operated with near-uniform efficiency
across the temperature range of our sample. While this may help to
reduce ICM abundances in groups, it cannot, on its own, explain the
SN~II trends in Fig.~\ref{fig,all}, and much less so for the SN~Ia
products. On the basis of the available data, however, we cannot
exclude the possibility that the discrepancy with respect to massive
clusters could contain a contribution from this effect.

\subsubsection{AGN-driven winds}\label{sec,agn}

The idea of central outflows driving material from group/cluster cores
has received much recent attention with the observation that AGN in
central cluster galaxies may be able to carve out cavities or bubbles
in the surrounding ICM through radio outbursts \citep{mcna07}.
Evidence of such activity has also been identified in small groups,
including several of the systems within our sample
\citep{alle06,mori06,jeth08,gast09}. An indication that some 
AGN--induced gas
mixing may have occurred in most, if not all, our groups is further
provided by the radial extent of the central abundance excess, which
is much broader than the distribution of optical light from the BGG
(cf.\ \citealt{rebu06}). Whether AGN-driven bubbles can rise far
enough in clusters to displace large amounts of material from cluster
cores remains unclear (e.g., \citealt{roed07}), but semi-analytic
models of galaxy formation have demonstrated that a variety of
observational constraints in the group regime can be reasonably well
satisfied in a scenario in which AGN may eject significant amounts of
hot gas from low-mass haloes \citep*{bowe08}. If so, AGN would provide
an attractive explanation for the gas and metal deficiency evident for
such systems in Fig.~\ref{fig,all}.

The standard assumption is that the observed cavities or bubbles have
arrived at their current location mainly by means of buoyancy
forces. If bubbles of enriched gas have travelled from group cores to
beyond $r_{500}$ by such forces only, the initial bubble entropy
$S_0=T/n_0^{2/3}$ must have been comparable to the ICM value $S_{500}$
at $r_{500}$.  We can evaluate the temperature difference
corresponding to $\Delta S = S_{500} - S_0$ and hence the energy
required to heat the relevant gas mass in group cores for it to be
buoyantly displaced beyond $r_{500}$.  In a similar vein to the
previous Section, we assume that the gas thus removed must account for
the lower $K$-band IMLR$_{\rm Ia}$ in the groups when compared to
NGC\,6338. To treat all groups equally, we assume $T_{500} = 0.67
\langle T\rangle$ for the ICM temperature at $r_{500}$ (cf.\ Paper~I,
equation~7), along with $Z=$~Z$_\odot$ for the removed gas as observed
in the group cores at present. For the initial entropy of the gas to
be displaced, we note that inspection of our {\em Chandra} images of
NGC\,741, HCG\,62, and NGC\,5044, all groups with known X-ray
cavities, suggests an average projected distance from cavity centre to
the optical centre of the BGG of roughly two-thirds of the $D_{25}$
semi-major axis. Given our sample mean of $\sim 22$~kpc for the
latter, this estimate is in excellent agreement with the mean
projected distance of $\sim 16$~kpc seen for the radio-filled cavities
in the combined group and cluster sample of \citet[see also
\citealt{raff06}]{birz04}. We therefore evaluate the initial gas
density and entropy at $r=0.7D_{25}$.

Results are plotted in Fig.~\ref{fig,agn}, in the form of the required
heating energy $\Delta E$. NGC\,4325 and NGC\,6338 itself are again
excluded, for the reason discussed earlier.  Most groups scatter
within a factor of a two in this plot, but NGC\,2300 and 4125 are
outliers due to their fairly high central entropies. The logarithmic
mean of the heating energy corresponds to $E\approx 8\times
10^{60}$~erg~s$^{-1}$, with no significant dependence on $\langle
T\rangle$ ($\sigma_K = -0.1$) or $L_{K,\rm BGG}$ ($\sigma_K =
+1.5$). While such energy requirements may exceed typical AGN outburst
energies inferred for local clusters (e.g., \citealt{raff06}), the
required energy could have been deposited in the course of multiple
outbursts and so may be broadly consistent with existing observational
estimates. We note, however, that the associated energy per total ICM
particle is generally somewhat in excess of typical values required by
preheating scenarios (mean of $\Delta E\approx
2.4$~keV~particle$^{-1}$), and does show some evidence for a negative
trend with $\langle T\rangle$ ($\sigma_K = -1.8$) but not with
$L_{K,\rm BGG}$ ($\sigma_K = 0.0$).

\begin{figure} 
 \includegraphics[width=84mm]{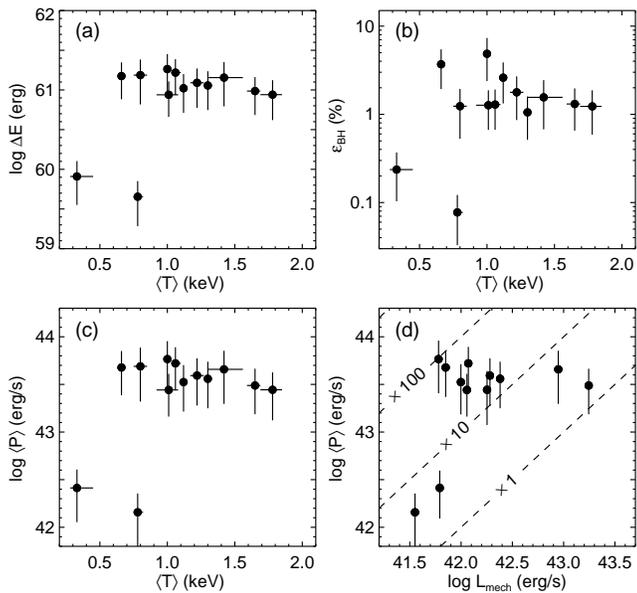}
 \caption{(a) Energy required to heat enriched gas in the group cores
  sufficiently for convection to carry this gas beyond $r_{500}$ and
  so achieve the same $M_{\rm Fe}/L_K$ ratio for SN~Ia products within
  $r_{500}$ as in NGC\,6338. (b) The corresponding required black hole
  energy conversion efficiency for the estimated BGG black hole
  masses. (c) The corresponding time-averaged heating power $\langle
  P\rangle$ assuming an energy injection time of 10~Gyr. (d) As (c),
  but compared to the estimated current mechanical power $L_{\rm
  mech}$ of the central AGN, with typical constant $\langle P\rangle /
  L_{\rm mech}$ ratios marked by dashed lines. Uncertainties reflect
  the fractional errors on IMLR$_{\rm Ia}$.}
\label{fig,agn} 
\end{figure} 

These results suggest that the energetic requirements for displacing
the required amount of gas and metals by buoyancy are generally large
for the current ICM configuration, but perhaps not prohibitively so.
Another way to arrive at this conclusion comes from considering the
implied efficiency $\epsilon$ with which the central supermassive
black hole (SMBH) must have converted accreted mass into outflow
energy, assuming that the released energy represents a converted rest
mass corresponding to the current mass of the SMBH. Estimating central
SMBH masses from the BGG central stellar velocity dispersions
$\sigma_{0, \ast}$ listed in Table~\ref{tab,BGG}, using
\begin{equation}
  M_{\rm BH} \sim 1.2 \times 10^8 
  \left(\frac{\sigma_{0,\ast}}{200\mbox{ km s$^{-1}$}} \right) ^{3.75}
\mbox{ M$_{\odot}$} 
\end{equation}
\citep{gebh00}, we show the corresponding conversion efficiencies
$\epsilon_{\rm BH} =\Delta E/M_{\rm BH} c^2$ in Fig.~\ref{fig,agn}b.
Results fall in the range $\epsilon \sim $\,0.1--5~per~cent, with a
median value of $\approx 1$~per~cent. Interestingly, but perhaps
coincidentally, this is in broad agreement with recent estimates of
the actual SMBH efficiency in converting accreted rest mass into
outflow energy, $\epsilon \approx 0.5$--2~per~cent (e.g.,
\citealt{alle06,merl08}).

Further insight can be gained from comparing the required
time-averaged heating power $\langle P\rangle$ to estimates of the
current mechanical AGN output in the groups. Fig.~\ref{fig,agn}c shows
the mean power associated with the derived $\Delta E$ for a
conservative assumption for the energy injection time-scale of 10~Gyr
(again corresponding to an initial redshift $z\approx 2$).
Fig.~\ref{fig,agn}d compares this to rough estimates of the currently
available power, based on the 1.4~GHz radio power $P_{1.4}$ of any
central radio source as listed in Table~\ref{tab,BGG}. We converted
$P_{1.4}$, where available, into current mechanical AGN power $L_{\rm
mech} \approx 10^{25} L_{\rm radio}^{0.44}$, assuming $L_{\rm radio}
\approx (8.6\times 10^{9}\mbox{ Hz}) P_{1.4}$ as is appropriate for a
power-law radio spectrum of index unity \citep{birz04}. Results reveal
that current central AGN activity in the groups is typically an order
of magnitude weaker than the inferred time-averaged power required to
accomplish the necessary gas removal. It is conceivable, however, that
this may at least partly be related to selection effects, since any
system with a powerful active radio source strongly disturbing the ICM
might not have made it into our sample, which only includes groups
which appear reasonably undisturbed on large scales in {\em Chandra}
data.

While acknowledging that the estimates in Fig.~\ref{fig,agn} result
from rather simplistic considerations which, among other things,
ignore the potential contribution from AGN not associated with the
BGG, the derived energy requirements suggest that some buoyant
displacement of gas to beyond $r_{500}$ by AGN activity may have taken
place in the recent past, but the majority of any such activity is
likely to have occurred when the ICM entropy configuration was more
conducive to blow-out. Also note that our energy estimates are based
on total mass losses assuming solar abundances for the removed gas,
but that larger amounts of gas (and hence lower metallicities and
larger outflow energies) would be needed if AGN were to also explain
the {\em general} shortfall of hot gas compared to clusters evidenced
by Fig.~\ref{fig,starbar}. In combination, this argues in favour of
gas and metal loss predominantly, but not necessarily exclusively, at
earlier epochs, closer to the redshift peak of metal production and
possibly prior to group collapse.

We do note, however, that recent AGN outflows could easily have
displaced highly enriched gas from group cores and so have contributed
to the central $Z$--$T$ trend in Fig.~\ref{fig,ZT}.  Using similar
assumptions as those underlying Fig.~\ref{fig,agn} but only
considering the `missing' Fe mass within $0.1r_{500}$ as derived from
the data in Fig.~\ref{fig,MFe_Lk}a, and assuming that this gas has
simply been expelled beyond $0.1r_{500}$, the resulting energy
requirements display sample mean values of $\langle \mbox{log\,}\Delta
E\rangle \approx 59.6$~erg, $\langle \eta_{\rm BH}\rangle \approx
0.1$~per~cent, $\langle \mbox{log\,}P\rangle \approx
42.1$~erg~s$^{-1}$, and $\langle (\mbox{log\,}P)/(\mbox{log\,}P_{\rm
mech}) \rangle \approx 1.0$. These requirements appear perfectly
reasonable, and suggest that current AGN power in the groups is
generally sufficient to buoyantly displace the required amount of
enriched gas from within $0.1r_{500}$. The requirements are generally
slightly harder to satisfy in hotter systems (as required by
Fig.\ref{fig,ZT}), with both required log\,$\Delta E$ and
$(\mbox{log\,}P)/(\mbox{log\,}P_{\rm mech})$ showing a weakly
significant tendency to increase with $T$, at $\sigma_K=+1.4$ and
$\sigma_K=+1.6$, respectively.

\section{Preheating and metal ejection from feeder filaments}
\label{sec,filaments}

The discussion in the preceding Section focussed on explaining the low
IMLR in groups in the framework of a decreased potential for {\em in
situ} metal production/release or retention in present-day group-sized
haloes. Motivated by the limited success of these explanations, with
the possible exception of AGN feedback, we now consider the
possibility that it is mainly related to processes occurring at the
epoch of galaxy formation. For example, the AGN results in
Section~\ref{sec,gasloss} demonstrate that substantial, although not
necessarily prohibitive, amounts of energy need to be injected into
the ICM to accomplish the necessary gas and metal removal at low
redshift from these groups. It would, however, have been energetically
favourable to remove some of this material at a much earlier stage,
when potential wells were shallower, and the gas more tenuous and
likely exhibiting a more irregular geometry much more conducive to gas
removal.  It seems attractive to connect this possibility with the
result that both star formation and AGN activity were much stronger at
high redshift, both peaking around $z \sim 2$--3
\citep*{bouw07,hopk07}. This section summarizes evidence that a
significant fraction of metals in groups must have been generated
before or during this phase, and discusses to what extent
proto-galactic winds may have heated and dispersed this material
beyond what is now $r_{500}$.

\subsection{The case for early enrichment}

There is substantial evidence that a significant fraction of the
present-day metals in the Universe must have been generated at very
early epochs. Enrichment of the intergalactic medium was taking place
already at $z\sim 6$ \citep{simc06}, with a non-negligible component
of the produced metals residing in Lyman break galaxies, some of which
presumably evolved into the massive cluster ellipticals seen today.
Contenders for this early enrichment include proto-galactic winds
occurring when (what are now) massive early-type galaxies formed the
bulk of their stellar mass. In galaxy clusters, evidence for
substantial enrichment prior to cluster collapse has been reported on
the basis of measured large-radius abundances \citep{fuji08}. More
generally, results for large cluster samples indicate that the hot gas
had achieved a mean Fe abundance of $Z_{\rm Fe} \approx 0.3$~Z$_\odot$
by $z\sim 1$, comparable to that seen in nearby clusters, although
some evolution in the central Fe content has occurred since then
\citep{bale07,maug08}.  The latter result draws support from the
simulations of \citet{cora08}, which suggest that cluster ICM
abundances at large radii ($\sim r_{500}$) were largely established at
$z\ga 1$, with only central abundances increasing significantly since
then. In smaller systems, the situation may be less clear
\citep{fino02} but in the specific case of our groups, a couple of
results do suggest that the material in their outskirts was
predominantly enriched at early epochs by core-collapse SNe, whereas
the abundance excesses in group cores may have been generated more
recently.

First, the inferred number ratio of SN~Ia to core-collapse SN in group
cores of $\approx 0.4$ is consistent with that measured in the local
($z\la 0.7$) Universe in the {\em Hubble} and {\em Chandra} Deep
Fields, but disagrees with predictions at $z\ga 1$ for reasonable
evolutionary models \citep{dahl04}. In the absence of serious recent
gas mixing (supported by the relatively undisturbed X-ray morphology
and the pronounced Fe gradients of our groups), this suggests that the
SN~Ia--dominated group cores in general, and the central Fe and Si
excesses in particular, have been generated within the last $\sim
$5--7~Gyr. The abundance pattern seen in the heavily SN~II--dominated
outskirts which represent most of the ICM is, on the other hand,
consistent with expectations for the SN~Ia/SN~II ratio at redshifts
$z\ga 1$ \citep{dahl04}. Hence, most of the SN~II activity in these
groups likely took place at high redshift.  This is in accordance with
the result that most of our groups show a galaxy morphology--density
relation of comparable strength to that observed in clusters
\citep{hels03}, suggesting that star formation and hence SN~II
activity has largely ceased at present in many of the bright galaxies
in the central regions.

Second, even when allowing for enrichment through stellar mass loss
from the central early-type in each group, the discussion in
Section~\ref{sec,SN} shows that an abundance contribution from SN~II
is, on average, required {\em everywhere} inside $r_{500}$.  The
near-uniform distribution of SN~II products revealed by
Fig.~\ref{fig,SN} suggests either an earlier enrichment epoch for
these than for the SN~Ia ejecta, allowing time for substantial mixing
before and during group collapse, or a much more efficient
redistribution mechanism for SN~II ejecta. However, the results of
\citet*{ferr00a} suggest that the distribution of SN~II products in
the absence of other mixing mechanisms should roughly follow that of
the galaxy light. While this is observed for the central SN~Ia
products in at least some of our groups \citep{mend09}, the SN~II
metals seem to trace more closely the much more extended distribution
of the hot gas. The idea that SN~II enrichment largely occurred before
group/cluster collapse, as advocated by, for example, \citet{fino01},
implies that the metals seen in group outskirts must have been
generated by SN~II before a significant number of SN~Ia began
exploding. Depending on the exact delay time distribution of SN~Ia,
this would imply SN~II iron enrichment to a level of $\sim
0.1$~Z$_\odot$ (Fig.~\ref{fig,SN}) at redshifts $z\ga 2-3$. As
discussed earlier, the gentle central rise seen for SN~II products can
then be explained if a relatively small fraction of these metals were
released after the group collapsed, possibly facilitated by
galaxy--galaxy and galaxy--ICM interactions, and with an additional
contribution associated with stellar wind loss from the central
early-type galaxy.

By considering the typical time-scales for metal production in the
groups, an additional argument can be provided in support of a
prominent high-redshift contribution to ICM enrichment. For this, we
focus first on the Fe produced by SN~Ia, which is less likely than
SN~II products to become locked up in stars in the early-type galaxies
that dominate the $K$-band output in our groups.  Using the same
general approach as for Fig.~\ref{fig,MFe_Lk}, while assuming a
constant SN~Ia rate per unit $L_K$ equivalent to the value observed in
local early-types \citep{mann05}, we plot the resulting enrichment
times corresponding to the total $K$-band IMLR$_{\rm Ia}$ within
$r_{500}$ in Fig.~\ref{fig,tot_times}a.  Under these assumptions, the
enrichment times should again be considered lower limits, because no
account is made of metals locked in stars, metals that have left the
group potential, or the growth of stellar mass within the group due to
galaxy assembly and the addition of galaxies to the group over the
time-scales involved. With results suggesting time-scales of $\ga
10$~Gyr in 40~per~cent of our groups, typically in the hotter and more
SN~Ia--dominated ones, it is far from clear that specific SN~Ia rates
seen in present-day early-types can account for the observed metal
masses. The discussion in previous Sections further indicates that
contributions from stellar mass loss and intracluster stars are
insufficient for bringing these time-scales below $\sim 10$~Gyr in all
groups; either the specific SN~Ia rate must increase with redshift, or
`external' enrichment must be invoked to explain the observed amount
of Fe from SN~Ia.

\begin{figure} 
 \includegraphics[width=80mm]{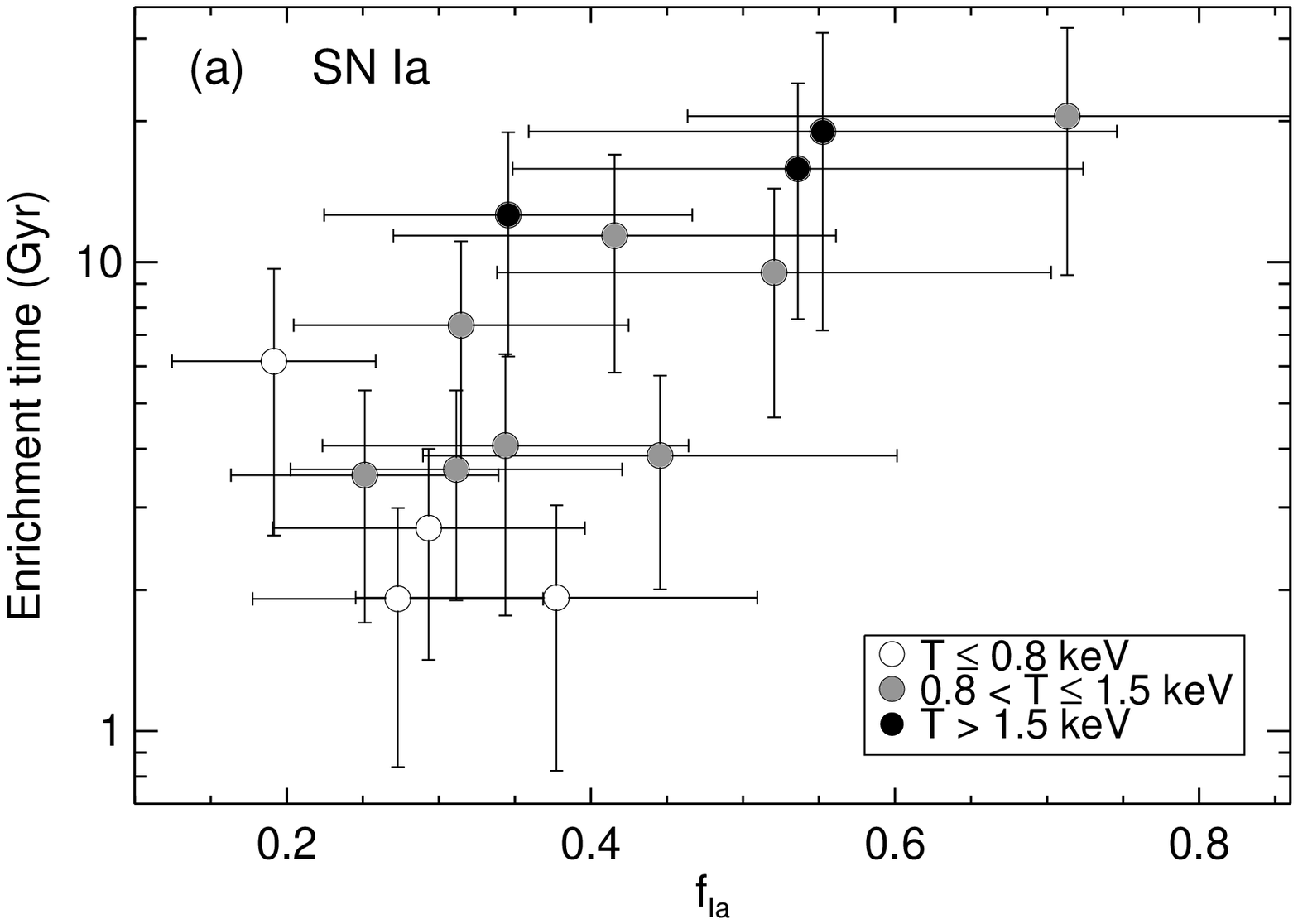}
 \includegraphics[width=80mm]{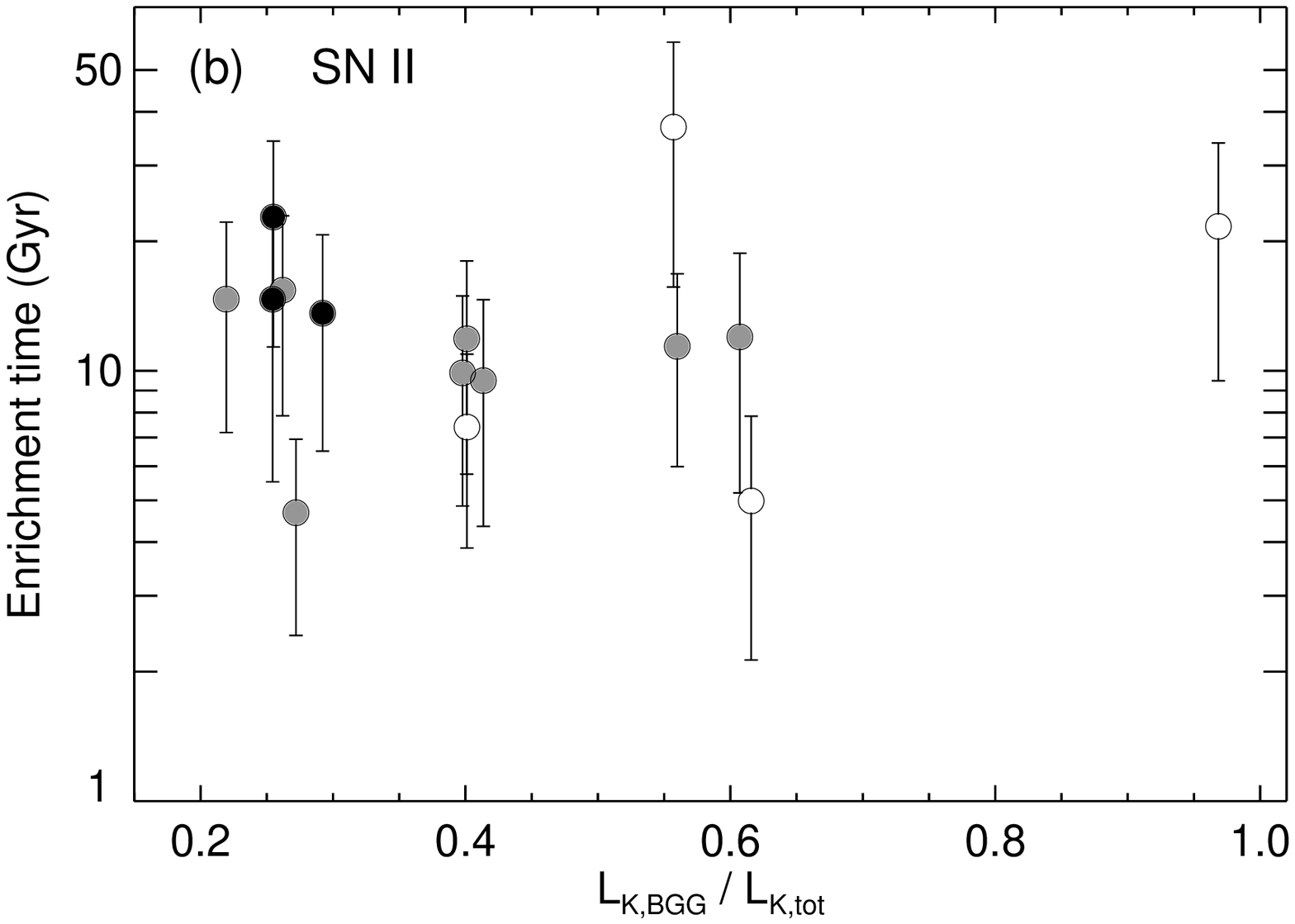}
 \caption{(a) SN~Ia enrichment times based on the $K$-band IMLR$_{\rm
    Ia}$, plotted against the fractional Fe mass from SN~Ia. (b)
    Similar for SN~II, assuming local Sbc/d SN~II rates for all
    satellite galaxies, as a function of the BGG contribution to total
    $K$-band light.}
\label{fig,tot_times} 
\end{figure}

This issue is even more acute for SN~II, whose contribution to Fe
enrichment is globally at least as important as that of SN~Ia (cf.\
Table~\ref{tab,gas}), but for which only upper limits to the observed
rate in local early-types are available \citep{mann05}. These
generally imply enrichment times far in excess of the age of the
Universe. However, some morphological transformations and truncation
of star formation has likely taken place in the group members
\citep{hels03}, so assuming current early-type SN~II rates could well
underestimate the metal output over cosmic time. In a crude attempt to
correct for this, we employ the conservative assumption that all
stellar mass not currently in the BGG has produced SN~II according to
the rate in local late-type {\em spirals}. The result is shown in
Fig.~\ref{fig,tot_times}b. As is clear, time-scales are still
prohibitively large in several cases, implying that SN~II rates must
also evolve with redshift, and even more so than for SN~Ia. Specific
star formation rates, and hence metal production rates, must have been
considerably higher in the past in these groups. Though not a
surprising result, it lends independent support to the notion that an
important fraction of ICM metals within $r_{500}$ was generated at
early epochs.

\subsection{Linking enrichment to preheating}

With much of the ICM pollution likely taking place around $z \sim
2$--3, close to the redshift peaks of the cosmic star formation rate
and of the AGN luminosity density, it is conceivable that metals and
energy were released from galaxies in substantial amounts before gas
and galaxies collapsed into groups, i.e.\ while still located in
filaments of low overdensities. This would not only suggest that the
trends in Fig.~\ref{fig,all} were at least partly established at this
epoch, but also offer a possible explanation of their origin in gas
and metal loss from the filaments that fed the growth of the group
structures observed today. Accretion along filaments implies, unlike
the case for virialized structures, a preferred direction along which
the ejection and loss of galactic material is strongly favoured. Given
the available data, we can only make some general speculations about
this possibility, but it seems an attractive one in light of the
somewhat limited success of the other explanations considered.

ICM enrichment at high redshift could possibly be linked to the
putative `preheating' invoked in some models, by which the ICM in
these systems is assumed to have had its entropy raised at high
redshift in order to explain the similarity breaking in the X-ray
scaling properties of groups and clusters.  Simulations and
observations (\citealt{borg08a} and references therein) indicate that
the associated energy required is of order $\sim 1$~keV per particle,
depending on the exact redshift of energy injection. It cannot greatly
exceed this value, as preheating would otherwise have destroyed the
ICM in $T\la 1$~keV groups. While the initial argument for introducing
this was to explain the seemingly elevated entropy levels in group
cores, preheating could also affect the gas at larger radii. For
example, the low gas mass-to-light ratio within $r_{500}$ in cool
systems could be related to the possibility that groups have seriously
inflated ICM distributions as a consequence of preheating through
early galactic winds. The resulting entropy of the heated (and
possibly enriched) gas would make it more resilient to gravitational
compression, to a certain extent preventing its re-collapse into the
group structures seen at low redshift. This is somewhat akin to the
`preferential infall' scenario proposed by \citet{fino01}, in which
small groups preferentially accrete low-entropy, low-metallicity gas
that was only mildly heated at high redshift.

One implication would be that a relatively larger fraction of any
gravitationally bound X-ray gas and metals should reside beyond
$r_{500}$ in cooler systems. We are not in an optimal position to test
this, as it would require sensitive X-ray data extending beyond this
radius for all groups. We can only note that with the adopted density
parametrizations, the ratio of gas mass within $0.5 r_{500}$ to that
within $0.5-1r_{500}$ actually shows a weak {\em anti}-correlation
with $\langle T\rangle$ ($\sigma_K = -1.6$) for our groups, consistent
with the correlation seen between $\beta$ and $T$ for the GEMS sample
in general ($\sigma=-1.4$; \citealt{osmo04}), from which the majority
of our groups were selected.  However, \cite{sand03} do find a
positive dependence of $\beta$ on $T$ when including clusters in the
comparison. The group--cluster discrepancy within $r_{500}$ could
contain some contribution from this effect, and if so would suggest
that a more fair comparison of systems across a wide range of masses
should be conducted within radii of lower overdensity such as
$r_{200}$.

Our speculation that a considerable fraction of the (centrally
concentrated) SN~Ia products were released after group collapse
suggests that at least part of the enrichment is not associated with
preheating but must have occurred at later times at $z\la 1$. This
agrees with the conclusion of \citet{fino02}, based on the absence of
a correlation between entropy at radii of equal enclosed mass vs.\ Fe
or Si abundance across a wide range of system masses. However, SN~II
provide the clear majority of the estimated SN energy released in our
groups, with an average contribution to $E_{\rm SN,\ast}$ in
Fig.~\ref{fig,release} of $\approx 0.8$~keV~per~particle, most of
which was presumably released in early starbursts prior to group
collapse. A potential caveat to the assumption that this value also
corresponds to the energy {\em delivered} to the ICM by galaxies is
that much of the SN explosion energy may have been rapidly radiated
away in the initial starburst phase rather than released from galaxies
in kinetic form. Whereas radiative losses are presumed to be minor at
low redshift, with direct escape of the SN explosion energy
facilitated by the presence of bubbles and chimneys in the ISM carved
out by previous generations of SN (and winds from their progenitors),
the ISM was denser, colder, and less porous at high redshift, possibly
enhancing radiative losses. If so, the $\approx 0.8$~keV~per~particle
inferred from the stellar mass of the group members may represent an
upper limit to the actual SN~II energy delivered to the galaxy
surroundings. On the other hand, simulations \citep{stri00} indicate
that coordinated SN explosions, such as those likely taking place
during vigorous high-redshift starbursts, tend to lower radiative
losses. Furthermore, Fig.~\ref{fig,release} suggests that the
estimated SN energy associated with the ICM metal mass broadly matches
the `available' energy, so any overestimation of the SN energy
imparted to the ICM should be modest.

Even if only a minor fraction of this energy were actually delivered
to the surroundings in proto-galactic winds, it could still have
contributed substantially to pre-heating the filaments feeding group
formation. It may not be sufficient to entirely explain the similarity
breaking between groups and clusters, or the presence of an `entropy
ramp' in group and cluster cores, but it certainly eases the energetic
requirements imposed on any other source to deliver the remaining
heating, and it may leave limited scope for AGN feedback. In the
cluster regime, numerical simulations \citep{torn04} suggest that a
feedback mechanism not associated with SN has to be acting at early
stages of cluster formation, before most of the mass is accreted in
the cluster potential well, in order to avoid the centrally peaked Fe
distribution seen in some simulations. At lower masses, however, where
our results show stronger Fe gradients that in clusters, the need for
AGN feedback is perhaps less pressing. The simulation work of
\cite{dave08}, for example, indicates that elevated entropy levels in
groups may be largely explained by radiative cooling combined with
SN-driven outflows, with no strong need for AGN-driven preheating.

Regardless of the heating source, the energetic requirements for
blow-out of gas via buoyancy discussed in Section~\ref{sec,gasloss}
would be strongly relaxed if the surrounding material is at a lower
overdensity than currently seen in group cores. For example,
Fig.~\ref{fig,agn} is based on assuming current ICM densities within
$\sim 10-20$~kpc of the group cores, which generally translate into
cosmic overdensities of order $\sim 10^3$. If the gas to be heated has
lower overdensities by a factor $\sim 100$ (certainly realistic for
pre-virialized structures such as inflowing filaments), the
requirements for heating it to current entropies at $r_{500}$ would be
reduced by a factor of $\sim 20$, i.e.\ to just $\sim
0.1$~keV~particle$^{-1}$ on average. This should be within range of
{\em either} SN or AGN at $z\sim 2-3$, suggesting that the energy
required to drive metals out of filaments is, indeed, available.

Whether the balance between bulk gas loss and selective loss of both
SN~Ia and SN~II products suggested by our results can be achieved is a
non-trivial question though. The shortfall of SN~II products in groups
seems largely explicable on the basis of pre-collapse metal loss in
starburst winds, but removal of large amounts of Fe from SN~Ia at
early epochs may also require some of this material to have been
synthesized very rapidly, for example by a population of SN~Ia
exploding promptly after star formation. The existence of such a
population has been proposed by \citet*{mann06} on the basis of
observed SN~Ia rates at high redshift, but the need for this
population at early times is still debated \citep*{dahl08}. 
Nevertheless, even in the `standard' SN~Ia delay-time
scenario, a non-negligible fraction of SN~Ia products may have been
released at fairly high redshift. For example, the model of
\citet{cora06} predicts a peak in the SN~Ia iron injection rate at
$z\sim 2-3$ in satellite galaxies and even earlier in the central
galaxy.

While some SN~Ia material {\em may} thus have been ejected from cool
systems in early starburst winds, our discussion of Fig.~\ref{fig,SN},
which shows that the SN~Ia contribution becomes insignificant outside
$\sim 0.7r_{500}$ still argues against a significant pre-collapse
SN~Ia contribution and so against substantial loss of SN~Ia metals
from filaments. To explore this issue further, we plot in
Fig.~\ref{fig,snia} Fe masses from SN~Ia within different radial
intervals, all simply normalised by total stellar light $L_K$. The
plot suggests that, over the mass range considered here, the reduction
in IMLR$_{\rm Ia}$ in lower-mass systems has preferentially occurred
within the central group regions, with no trend with $\langle
T\rangle$ seen beyond $0.5r_{500}$. This preferential loss of SN~Ia
products from the central Fe peak seems consistent with post-collapse
AGN outflows (and possibly some cooling) as important contributors to
the shortfall of SN~Ia products in cool systems. The considerations in
Section~\ref{sec,agn} then suggest that this loss cannot generally
have occurred very recently but rather must have taken place over
considerable time-scales following group collapse.

\begin{figure} 
 \includegraphics[width=84mm]{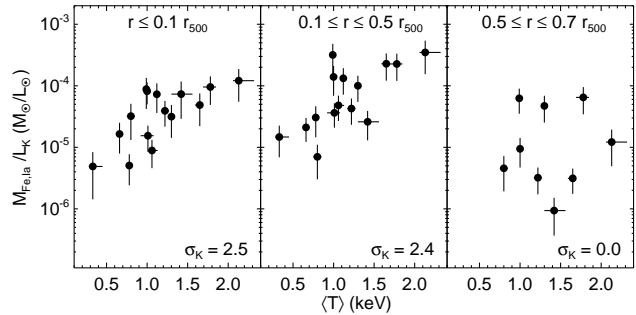}
 \caption{Mass of Fe from SN~Ia within different radial ranges,
   normalised by total $L_K$. As in Fig.~\ref{fig,ZT}, only groups
   with data covering the full radial interval are shown in each
   panel. Labels in lower right corners specify the significance
   $\sigma_K$ of any linear correlation.}
\label{fig,snia} 
\end{figure}

\section{Discussion and conclusions}\label{sec,conclude} 

Based on the results presented in Paper~I in this series
\citep{rasm07}, in this second of two papers we have investigated the
distribution and total mass of SN~Ia and SN~II products in the ICM of
15 groups, for comparison to the total ICM mass and to the optical
luminosities and stellar masses of the group members. We re-emphasize
here that our group sample was not intended to be statistically
representative in any sense. Nevertheless, the results should still
provide useful constraints on models of the evolution of galactic
systems on scales from small groups to massive clusters. We note that
there are features of the observed abundance distributions that we
have not covered in any detail. These include the broad extent of the
central abundance excess, reaching well beyond the distribution of
stellar light from the central galaxy, as well as the curious
abundance dips seen for both Fe and Si at the very centre for several
groups, manifest also in the averaged profiles in
Fig.~\ref{fig,norm_bin}.  Any detailed discussion of these features is
likely to benefit from comparison of group properties to those of
non-cool core systems, however, and so is deferred to a future paper.
Our existing results can be summarized as follows:

\begin{enumerate}

\item[1.] As anticipated, the total mass of iron and silicon in the
  hot gas within $r_{500}$ scales strongly with total group mass as
  measured by the mean X-ray temperature outside any cool core. Values
  range from $M_{\rm Fe}=(0.4-23) \times 10^8$~M$_\odot$ and $M_{\rm
  Si}=(0.6-35) \times 10^8$~M$_\odot$ across the 0.3--2.1~keV
  temperature range of the sample. SN~II have provided on average
  $\sim 60$~per~cent of the Fe and $\sim 90$~per~cent of the Si inside
  $r_{500}$, with a mildly significant tendency (at $\leq 2\sigma$)
  towards increasing SN~II dominance in cooler systems.

\item[2.] For the sample as a whole, the distribution of SN~Ia
  products is strongly peaked towards the group core, whereas SN~II
  enrichment is nearly uniform throughout the ICM. The Fe mass within
  $0.1r_{500}$ (roughly the size of the cool core, where present) can
  be explained by prolonged enrichment by SN~Ia in the central
  early-type galaxy, aided by stellar mass loss and intracluster
  stars. In particular, if adopting a plausible model for the cosmic
  evolution of the SN~Ia rate, typical enrichment time-scales are $\la
  5$--7~Gyr in all groups, comparable to similar results for central
  cluster galaxies. The mild central rise observed for SN~II products
  can be explained by stellar mass loss from the central galaxy,
  potentially with an additional contribution from metals released by
  other group galaxies via density--dependent release mechanisms such
  as galaxy--ICM interactions.
  
\item[3.] Due to the combination of lower gas mass-to-light ratios and
  metal abundances in cooler systems, resulting aggregate metal
  mass-to-light ratios within $r_{500}$ show an order-of-magnitude
  variation across the sample, only approaching results for massive
  clusters at the high--temperature end. For a given amount of stellar
  light, there is thus a substantial shortfall of X-ray metals in
  cooler systems, a deficiency which is found to be present both for
  SN~Ia and SN~II products. This suggests that metals from both
  SN~types were either synthesized or released from galaxies less
  efficiently in lower-mass systems, that metals were lost from such
  systems, or that the intergalactic gas accreted by small groups was
  less enriched than that accreted by more massive systems

\item[4.] Both Fe and Si abundances in group cores ($r\la 0.3r_{500}$)
  show a positive trend with group temperature which is significant at
  $\sim 3\sigma$. No corresponding trend is seen at larger radii
  ($r\ga 0.5r_{500}$) however, arguing against recent preferential
  loss of enriched material from cooler systems as an explanation of
  their lower global metal-mass-to-light ratios. Estimates of the
  current mechanical power of the central AGN in the groups indicate
  that the metallicity trend in group cores may at least be partly due
  to the expulsion of enriched material to beyond $0.1r_{500}$ by
  AGN--driven buoyant bubbles, perhaps with some contribution from
  radiative cooling of enriched gas in the core, causing less enriched
  material to flow inwards. Alternatively, it can be understood as due
  to the action of a metal release mechanism whose efficiency depends
  on local ICM density and the total mass of the system, such as ram
  pressure stripping, but it cannot be explained by a systematic
  variation in the contribution from intracluster stars with group
  temperature. The absence of a similar reported trend in systems
  above $T \approx 2$~keV may indicate that the relevant processes
  `saturate' in cluster cores, perhaps as a consequence of
  near-complete gas removal from core galaxies in such environments,
  or due to the larger heating energies or time-scales required to
  accomplish significant metal removal by AGN outflows or radiative
  cooling in such systems.

\item[5.] As a consequence of the lower Fe abundances and mildly
  increased contribution from SN~II in cooler systems (i.e.\ higher
  $\alpha$/Fe~ratios), we find tentative evidence for the ICM within
  the central few hundred kpc ($r\la 0.5r_{500}$) to appear chemically
  less `evolved' in these groups. This may reflect a more extended
  star formation history in less massive groups, consistent with the
  slightly bluer average galaxy colour seen in these systems.
  
\item[6.] Metal release from galaxies to the ICM appears to have been
  efficient in these groups, with, on average, an estimated $\sim
  70$~per~cent of the metals synthesized in galaxies currently present
  in the ICM. While this is in fair agreement with recent independent
  estimates in clusters \citep{siva09}, there is also some evidence
  for the release fraction to be systematically lower, and well below
  cluster values, in the cooler groups in our sample.

\item[7.] Among possible explanations for the metal deficiency in
  groups within $r_{500}$, we can rule out radiative cooling as an
  important gas and metal sink. Some contribution may instead come
  from the more extended star formation histories and lower metal
  release efficiencies in cooler systems, as well as the tendency for
  such systems to contain a relatively larger fraction of their gas
  beyond $r_{500}$. Within the current ICM configuration, the
  energetic requirements for the necessary gas and metal loss induced
  by AGN-- or SN-driven winds are rather large (although not
  necessarily prohibitively so), generally favouring loss of material
  at the incipient stages of group evolution. Other mechanisms such as
  thermal conduction transporting gas and metals out beyond $r_{500}$
  in cooler systems can probably be ruled out, due to the strong
  temperature dependence anticipated for heat conduction unsuppressed
  by magnetic fields (Spitzer conductivity $\propto T^{5/2}$).

\item[8.] Comparison to observed ratios of SN~Ia to core-collapse SN
  in deep field data \citep{dahl04} shows that group cores display
  abundance ratios consistent with inferred SN ratios in the
  low-redshift Universe ($z\la 0.7$) but are inconsistent with
  predictions for $z\ga 1$. This argues in favour of relatively recent
  enrichment of the central group regions, including the build-up of
  the central abundance excesses over the past $\la 6$--8~Gyr. In
  contrast, the heavily SN~II-dominated abundance ratios in group
  outskirts does agree with predictions for $z\ga 1$, suggesting that
  most of the ICM was enriched to $\sim 0.1$ solar at redshifts close
  to the peak of the cosmic star formation rate. The near-uniform
  radial distribution of SN~II products corroborates this picture by
  arguing in favour of substantial SN~II enrichment prior to or during
  group collapse.

\item[9.] This raises the possibility that powerful starburst winds at
  high redshift have contributed significantly to the gas and metal
  deficiency in groups, causing loss of baryons from the filaments
  feeding the growth of these structures. The estimated SN explosion
  energy associated with high-$z$ star formation in the group members
  seems sufficient for this, and may further have provided a
  substantial fraction of the preheating energy required by some
  models for explaining the X-ray scaling properties of groups. If so,
  it is not clear that a prominent contribution to preheating from
  quasar activity is required in these systems. Post-collapse AGN
  feedback can nevertheless still have played a prominent role in
  subsequently ejecting mainly SN~Ia metals from group potentials.

\end{enumerate}

One implication of our results is that present-day clusters cannot
have formed through hierarchical build-up of smaller structures
resembling the groups studied here. This is, of course, already clear
from the well-established differences in cold fraction (cf.\
Section~\ref{sec,SFE}), but the deficiency of gas-phase metals in
smaller systems poses an additional problem for this scenario.  Of
course there is no compelling reason to believe that present-day
groups are chemically similar to those that participated in cluster
build-up at higher redshift, although our considerations suggest that
the metal deficiency in low-mass systems has not been established
recently. Alternatively, the proto-clusters into which such groups
merged have been massive enough to retain or re-accrete enriched
material shed by their smaller progenitors.  Deep X-ray observations
of high--$z$ groups may eventually shed light on these issues, but the
questions are unlikely to be settled by the current generation of
X-ray satellites.  Numerical work may also provide clues to the
chemical properties of cluster progenitors and in which ways these
depart from those of our groups.  A prerequisite for such a
comparison, however, is that models can accurately reproduce
observational results for the ICM abundance distribution in groups at
low redshift.  While significant improvements have been made in the
past few years in this regard, some issues remain such as reproducing
the observed distribution of $\alpha$--elements (see, e.g.,
\citealt{dave08}).

Alternatively, our groups are conceivably not representative of
(X-ray) groups in general. Although exhibiting X-ray luminosities that
span two orders of magnitude \citep{osmo04}, our group sample is still
largely restricted to cool-core groups that appear morphologically
undisturbed. In order to gain a more complete picture of the
enrichment and feedback history in groups, similar analyses would need
to be extended to a more heterogeneous group sample, including a
substantial number of systems without a cool core. At higher masses,
such systems tend to show much flatter abundance gradients than those
reported here (e.g., \citealt{degr04}), suggesting fundamental
differences in ICM enrichment and heating histories between cool-core
and non--cool core systems. Such a comparison is the subject of
ongoing work (Johnson et~al., in preparation), and may ultimately help
in providing an improved understanding of the chemical and
thermodynamical evolution of baryons in both galaxies and the
intergalactic medium.

\section*{Acknowledgments} 
 
We are grateful to the referee for helpful and constructive comments
which helped to clarify the robustness of our conclusions. We
acknowledge useful discussions with Alexis Finoguenov, Alastair
Sanderson, and Fran\c{c}ois Schweizer. This work has made use of the
NASA/IPAC Extragalactic Database (NED) and the Two-Micron All-Sky
Survey (2MASS) database. JR acknowledges support provided by the
National Aeronautics and Space Administration through Chandra
Postdoctoral Fellowship Award Number PF7-80050 issued by the Chandra
X-ray Observatory Center, which is operated by the Smithsonian
Astrophysical Observatory for and on behalf of the National
Aeronautics and Space Administration under contract NAS8-03060.

\bsp

\label{lastpage} 
 
\end{document}